\documentclass[%
preprint,
 amsmath,amssymb,
 aps,
]{revtex4-1}

\usepackage{graphicx}
\usepackage{dcolumn}
\usepackage{bm}
\usepackage{color}
\usepackage{appendix}
\newcommand{\argmax}{\mathop{\rm arg~max}\limits}
\newcommand{\argmin}{\mathop{\rm arg~min}\limits}     

\begin{document}

\title{Appropriate basis selection based on Bayesian inference for analyzing measured data reflecting photoelectron wave interference}

\author{Yasuhiko IGARASHI}
\affiliation{Graduate School of System and Information Engineering, University of Tsukuba, 
  Tsukuba, Ibaraki 305-8577 , Japan}
\affiliation{Japan Science and Technology Agency, PRESTO,  Kawaguchi, Saitama, 332-0012, Japan}

\author{Fabio IESARI}
\affiliation{Kyushu Synchrotron Light Research Center, 8-7 Yayoigaoka, Tosu, Saga 841-0005, Japan}

\author{Hiroyuki SETOYAMA}
\affiliation{Kyushu Synchrotron Light Research Center, 8-7 Yayoigaoka, Tosu, Saga 841-0005, Japan}

\author{Toshihiro OKAJIMA}
\affiliation{Kyushu Synchrotron Light Research Center, 8-7 Yayoigaoka, Tosu, Saga 841-0005, Japan}
\affiliation{Aichi Synchrotron Radiation Center, 250-3 Minamiyamaguchi-cho, Seto-shi, Aichi-ken, 489-0965 Japan}

\author{Hiroyuki KUMAZOE}
\affiliation{Institute of Pulsed Power Science, Kumamoto University, Kumamoto, 860-8555, Japan}

\author{Ichiro AKAI}
\affiliation{Institute of Pulsed Power Science, Kumamoto University, Kumamoto, 860-8555, Japan}
\affiliation{Kyushu Synchrotron Light Research Center, 8-7 Yayoigaoka, Tosu, Saga 841-0005, Japan}

\author{Masato OKADA}
\affiliation{Graduate School of Frontier Science, The University of Tokyo, Kashiwa, Chiba, 277-8561, Japan}
\email{igayasu1219@cs.tsukuba.ac.jp}

\date{\today}
  

\begin{abstract}
In this study, we applied Bayesian inference to extended X-ray absorption fine structure (EXAFS) to select an appropriate basis from Fourier, windowed Fourier and advanced Fourier bases, and we extracted magnitude spectra obtained by these bases, which are closely related to a local structure near the target atom. Based on our method, we also estimate optimal physical parameters from EXAFS signals alone using prior physical knowledge, which can generally be realized in condensed systems. To evaluate our method, the well-known EXAFS spectrum of copper was used for the EXAFS data analysis. We found that the advanced Fourier basis is an appropriate basis for the quantitative regression of the EXAFS signal and that the estimation of the Debye-Waller factor can be robustly realized by using the advanced Fourier basis alone. Bayesian inference based on minimal restrictions allows us to not only eliminate some unphysical results but also select an appropriate basis. 
Bayesian inference enables us to simultaneously select an appropriate basis and optimized physical parameters without FEFF analysis, which allows extraction of magnitude spectra, which are closely related to a local structure, from EXAFS signals alone. These advantages lead to the general usage of Bayesian inference for EXAFS.
\end{abstract}

                             

\maketitle   

\section{\label{sec:intro}Introduction}
\begin{figure*}[t!]
  \centering
  \includegraphics[width=6in]{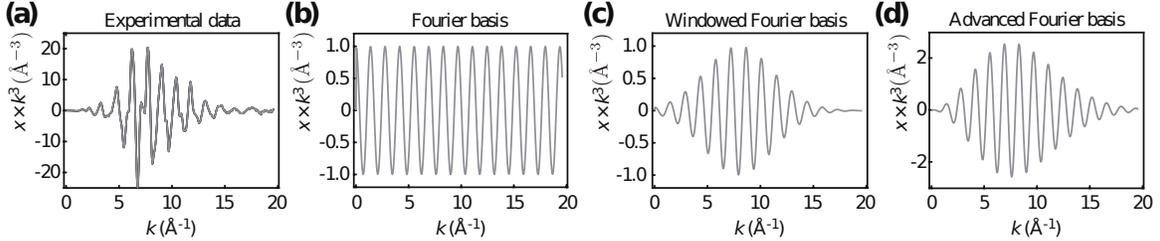}
  \caption{(a) Measured background-subtracted and $k^3$-weighted EXAFS data from copper foil measured at $T=300~K$. (b) Fourier basis. (c) Windowed Fourier basis. (d) Advanced Fourier basis.}
  \label{fig:Exp_FT_WT_AdvFT}
\end{figure*}

X-ray absorption fine structure (XAFS) measurements provide information about atomic-scale local structures and the local electronic states of various materials \cite{Bunker2010}. 
In this study, we focused on the analysis of extended X-ray absorption fine structure (EXAFS) measurements, 
as shown in Fig. \ref{fig:Exp_FT_WT_AdvFT}(a), 
in which EXAFS oscillations extend from $50$ to $1000$ eV above the absorption edge. These oscillations are caused by the interference of photoelectron waves emitted from an absorbing atom and backscattered by neighboring atoms. Based on this effect, an EXAFS spectrum can provide the coordination number of neighboring atoms as a function of the distance from the absorbing atom. 

Regarding the analysis of EXAFS data, software tools such as Athena \cite{Ravel2005}, which removes the baseline from the XAFS spectrum and performs a Fourier transformation (FT) of the extracted EXAFS oscillations, are frequently used 
and FT uses a Fourier basis of infinitely expanded periodic oscillations, as shown in Fig. \ref{fig:Exp_FT_WT_AdvFT}(b)
A Fourier transform of the EXAFS with respect to $k$ has peaks near the first few near-neighbor distances. 
Thus, EXAFS is used as a general probe of microscopic structural information in molecules and solids \cite{sayers1971new,kuzmin2000dehydration,Teo2012}. 
However, the magnitude of FT is thus substantially modified by the decoherence, phase shift, damping effects of photoelectron waves, 
and multiple-scattering effects from local structure \cite{Teo2012}. 

If one replaces the infinitely expanded periodic oscillations (Fourier basis) with located wave trains as the basis for the integral transformation, 
then one may analyze EXAFS data without information distortion of the coordination number of neighboring atoms and their distance from the absorbing atom. Previous studies have proposed two types of alternative bases. The first type is the wavelet transform basis \cite{funke2005wavelet,timoshenko2009wavelet,timoshenko2014exafs,timoshenko2018neural} 
and the Morlet wavelet \cite{grossmann1990reading} is often selected as a wavelet basis in EXAFS analysis \cite{funke2005wavelet}.  
In this study, we used a windowed FT (WindowedFT) basis, which is similar to the wavelet basis. This basis is obtained by taking a complex sine wave (as in a FT) and confining it with a Gaussian envelope, as shown in Fig. \ref{fig:Exp_FT_WT_AdvFT} (c). 

The second type is the advanced Fourier transformation (AdvFT) basis, which is derived from a simplified model based on the single-scattering approximation \cite{Akai2018}, as shown in Fig. \ref{fig:Exp_FT_WT_AdvFT}(d). 
The AdvFT basis looks quite similar to that proposed by Funke et al~\cite{funke2007new}. 
Although AdvFT and WindowedFT bases both have located wave trains similar to an EXAFS signal, 
the localized wave in AdvFT is asymmetrically shaped, which originates from the physical model of the single-scattering approximation. 
Using the AdvFT basis and the physical assumption that the atomic arrangement in a substance is actually discrete (or sparse), previous research has estimated not only 

the discrete magnitude spectra obtained by Advanced Fourier basis 
but also the Debye-Waller factor from experimental data alone \cite{Akai2018}. 
Despite these advantages, the WindowedFT and AdvFT bases have been used only occasionally for EXAFS data analysis,
and which basis is appropriate for the analysis of EXAFS signals remains unknown. 
This problem is called the basis selection problem, and it is one of the model selection problems \cite{bishop2006pattern}. 
In the basis selection problems, Bayesian inference is an effective statistics framework appropriately using both an prior knowledge and EXAFS signals \cite{bishop2006pattern}. 
In this study, we focus on the basis selection problem to show the effectiveness of Bayesian inference for EXAFS analysis. 

To analyze the EXAFS signals, a theoretical methods, and FEFF have been frequently used. However, to estimate the Debye-Waller factor, the prior knowledge of the structure is essential required. On the other hand, approaches based on the multicomponent/multishell fittings \cite{Bunker2010,martini2017composition}, the reverse Monte Carlo (RMC)\cite{iesari2018gnxas,timoshenko2012reverse} and molecular dynamics \cite{palmer1996direct,timoshenko2018probing} have also actively employed. The first two of them are fitting techniques to estimate the Debye-Waller factor and the RMC and MD provide fruitful information of three dimensional structures. The calculation cost for such approaches might be high to evaluate statistical accuracy of the estimated physical quantities such as radial distribution function. There is another approach to evaluate such accuracy only from one measured EXAFS signal. That is the Bayesian approach.

To avoid the situation and introduce an appropriate prior knowledge, as has been explained above, 
Bayesian inference is an effective statistics framework \cite{bishop2006pattern} since Bayesian inference enables us to simultaneously select an appropriate prior knowledge depending on the EXAFS signals. 
Regularization technique is often used in EXAFS analysis \cite{ershov1981new}. This can be considered as an analysis technique that incorporates this prior knowledge by using the Bayesian inference \cite{bishop2006pattern}. To extract the resulting model weights tend to be smaller in absolute value, the previous work introduced regularization terms. Then, the extracted local structure has lower variance and higher bias and this leads to avoid over-fitting on training data \cite{ershov1981new}. 
In this study, to include the sparsity of an atomic arrangement in a substance, we use L1-regularization \cite{Tibshirani1996} and 
optimize the regularization parameter, which is decided by hand or a practical way at previous research \cite{ershov1981new,kunicke2005efficient}. 
Moreover, recently a method based on Machine Learning including neural networks (NN) \cite{timoshenko2018neural} was successfully applied to the analysis of EXAFS, although this requires a specialized trained NN for each case and an Bayesian neural network will be needed for selecting an appropriate prior knowledge and evaluating the reliability of obtained results \cite{hafner2018reliable}. 

To examine the effectiveness of the Bayesian inference, we focus on the basis selection problem from the three basis, FT, WindowedFT and AdvFT bases in this study. Bayesian inference combines prior knowledge and a model of the observation process, as shown by previous studies for EXAFS signals \cite{Krappe2000,Krappe2002}. 

For deciding the atomic arrangement in a substance and physical parameters from EXAFS signals alone, 
we extend the previous works to select an appropriate basis and 
optimize the regularization parameter, which is decided by hand or a practical way at previous research \cite{kunicke2005efficient,ershov1981new}. 
We assumed the physical prior knowledge that is generally realized condensed systems, such as the observation process in EXAFS signals, the sparsity of an atomic arrangement in a substance, and the increase in the number of neighboring atoms with increasing atomic distance. To evaluate our method, the well-known EXAFS spectrum of copper and iron were used for the EXAFS data analysis. 

\section{Experimental data and advanced Fourier basis  based on physical model}
\subsection{Experimental EXAFS data}
To evaluate the validity of our proposed method, 
the well-known EXAFS spectrum of copper, which was 
measured by using the BL11 beamline at the SAGA Light Source \cite{Okajima2013}, was used as the object of analysis in this paper. 

An EXAFS signal $\chi(k)$ is an oscillating part of the X-ray absorption coefficient arising from the interference of the photoelectron waves between outgoing and backscattered waves \cite{Bunker2010,Calvin2013}. 
In $\chi(k)$, $k$ represents the photoelectron wavenumber, which is commonly converted from the X-ray energy $E$ as follows:
$k =\sqrt{2m(E - E_0)/\hbar^2}$. 
Here, $E_0$ is the absorption edge energy, $m$ is the electron mass, and $\hbar$ is Planck's constant. 

Following the prevailing custom, to emphasize oscillations at large $k$, the EXAFS oscillations are presented with multiplication of the $k^3$ term, as shown in Fig.~\ref{fig:Exp_FT_WT_AdvFT}(a). 
We then define $y(k)\equiv \chi(k)k^3$ for the application of our proposed method, where the abscissa is the wavenumber $k$, and its measurement range and the number of data points are $0\sim20$~\AA$^{-1}$~and $p=393$, respectively. 
The range of the interatomic distance $R_j$ from the X-ray-absorbing atom are $0\sim10$~\AA~and the total number of $R_j$ is set as $M=326$. Each point is placed at even intervals and we have divided the range of possible interatomic distances ($10$~\AA~) in uniform intervals, and the $R_j$ corresponds to the center of each of these intervals as follows: $R_j =\frac{10}{2M}+\frac{10(j-1)}{M}$~\AA~ $(j=1,2,\dots,M=326)$. 

\subsection{The single-scattering approximation for EXAFS signals}
A single-scattering curved-wave harmonic approximation is
widely used to describe a contribution from neighbor $N$ atoms 
located at a close distance $R$ from the absorber. In this case, 
$y(k)$, which is equal to the EXAFS signal, can then be modeled as follows: 
\begin{eqnarray}
y(k) = S _ { 0 } ^ { 2 } \sum _ { j } N \left( R _ { j } \right) t _ { j } ( k ) \frac { k ^ { 2 } } { R _ { j } ^ { 2 } } 
\times \exp \left\{ - 2 \left[ k ^ { 2 } \sigma _ { j } ^ { 2 } + \frac { R _ { j } } { \Lambda ( k ) } \right] \right\} 
{\sin \left[ 2 k R _ { j } + \delta _ { j } ( k ) \right],  }
\label{eq:SingleScatteringApprox}
\end{eqnarray} 

where 
$N(R_j)$ represents the coordination number of neighboring atoms at distance $R_j$. 

$S_0^2$ is an amplitude reduction factor \cite{Calvin2013} that must be determined independently to discuss the absolute values of $N(R_j)$. 
$t_j(k)$, $\Lambda(k)$, and $\delta_j(k)$ are the backscattering amplitude, the mean free path, and the phase shift as functions of the photoelectron wavenumber $k$, respectively. 

The $k^2$ term originates from the multiplier $k^3$ in $y(k)$ and leads to an increasing feature of oscillating amplitude in EXAFS signals. 
The $\exp(-2k^2\sigma_j^2)$ term exhibits decoherence effects and leads to a decreasing feature of oscillating amplitude at large $k$. 
Thus, these two factors result in asymmetric bell-shaped wave trains, as shown in Fig. \ref{fig:Exp_FT_WT_AdvFT}(a). 
As stated in the Introduction, it is necessary to incorporate these terms, which lead to a $k$-dependent amplitude in EXAFS oscillations, into basis functions instead of the Fourier basis functions, $\sin[2kR_j +\delta_j(k)]$, such as WindowedFT basis and AdvFT basis, as becomes apparent below. 

\subsection{Bases for EXAFS analysis}
In this section, let us consider three bases for EXAFS analysis: FT, WindowedFT and AdvFT bases. 
We first introduce the AdvFT basis, which is based on the single-scattering approximation for EXAFS oscillations. 
To surpass the analysis by the conventional FT basis, 
we derived a simplified physical basis function from the single-scattering processes, which fulfills the conditions of a linear system, as follows \cite{Akai2018}: 
\begin{eqnarray}
y(k) = \sum _ { j } \frac { k ^ { 2 } } { R _ { j } ^ { 2 } }
\exp \left\{ - 2 \left[ k ^ { 2 } \sigma_{\mathrm{DW}}^ { 2 } + \frac { R _ { j } } { \Lambda} \right] \right\} 
\times 
\left[ a_j\sin 2 k R_{j} + b_j\cos 2k R_{j} \right]. 
\label{eq:SingleScatteringConstApprox}
\end{eqnarray} 
Although Eq. (\ref{eq:SingleScatteringApprox}) includes the $R_j$ dependence of
the Debye-Waller factor and the $k$ dependence of the mean free path, 
these parameters in Eq. (\ref{eq:SingleScatteringConstApprox}) are simplified to be constant in $\sigma_{\mathrm{DW}}$ and $\Lambda$, respectively. 
In the proposed model, 
the sine function in Eq. (\ref{eq:SingleScatteringApprox}) is extended to sine and cosine functions with expansion coefficients $a_j$ and $b_j$, respectively, in the same manner as the FT. 
The expansion coefficients in this model absorb both the effect of the backscattering amplitude $t_j(k)$ and that of $S_0$. 
These parameters are summarized in Table \ref{tab:EXAFS_parameters}. 
 \begin{table*}[t!]
    \caption{Parameters for EXAFS analysis}
    \centering 
    \begin{tabular}{| l | c | c |} 
    \hline\hline 
                 & Single-scattering &                  \\ 
    Description & approximation   &   Advanced Fourier   \\ 
    \hline 
     photoelectron wavenumber & $k$ & $k$ \\
    interatomic distance from X-ray-absorbing atom & $R_j$ & $R_j$ \\
EXAFS signal &  $\chi(k)$ & $\chi(k)$ \\
Coordination number &  $N(R_j)$  & Included in coeff. $a_j$, $b_j$ \\
    Debye-Waller factor       &  $\sigma_j$ & $\sigma_{\mathrm{DW}}$(const.)\\
    amplitude reduction factor & $S_0^2$& Including in coefficients $a$, $b$\\
    backscattering amplitude & $t_j(k)$ & Including in coefficients $a$, $b$\\
    mean free path & $\Lambda(k)$ & $\Lambda$ (const.) \\
    phase shift   & $\delta_j(k)$ & 
    Including in coefficients $a$, $b$ \\
    \hline
    \end{tabular}
    \label{tab:EXAFS_parameters}
\end{table*}
\vspace{2mm}

These simplifications enable us to obtain a simplified physical basis function that satisfies both the linear system requirement and an atomic-scale local structure based on the single-scattering approximation. 
In this study, we call the basis the AdvFT basis function \cite{Akai2018}, 
and we define the basis function matrix $\boldsymbol{X}^{\mathrm{AdvFT}}$ 
as follows: 
\begin{eqnarray}
\boldsymbol{y} = \boldsymbol{X}^{\mathrm{AdvFT}}\boldsymbol{w} + \boldsymbol{\epsilon}. 
\label{eq:chik3=Ax}
\end{eqnarray}
We define ${\bf X}^{\mathrm{AdvFT}} = \{X_{i, j, l}^{\mathrm{AdvFT}}\} \;\;(\forall {i, j, l}, \{i=1,\dots,p=393\}, \{j=1,\dots, M=326\}, \{l=0, 1\})$ as follows: 
\begin{eqnarray}
  \boldsymbol{X}^{\mathrm{AdvFT}} = \left(
    \begin{array}{cccccc}
      1         & X_{1,1,0} & X_{1,1,1} & X_{1,2,0}   & X_{1,2,1} &  \ldots \\
      \vdots & X_{2,1,0}  & X_{2,1,1} & X_{2,2,0}  & X_{2,2,1} &  \ldots \\
      \vdots  & \ldots            & \ldots            & \ldots           & \ldots            & \ldots \\
      1         & \ldots            & \ldots            & \ldots           & \ldots            & X_{p, M, 1}, 
    \end{array}
  \right)\nonumber
\end{eqnarray}
with each element of the matrix ${\bf X}^{\mathrm{AdvFT}}$ being a representation of 
\begin{eqnarray}
X_{i, j, 0}^{\mathrm{AdvFT}}=\frac{k_i^2}{R_j^2} \exp\left[-2\left(k_i^2\sigma_{\mathrm{DW}}^2 + \frac{R_j}{\Lambda}\right)\right] \sin 2k_iR_j,
							\label{eq:ASincoeff}
							\\
X_{i, j, 1}^{\mathrm{AdvFT}}=\frac{k_i^2}{R_j^2} \exp\left[-2\left(k_i^2\sigma_{\mathrm{DW}}^2 + \frac{R_j}{\Lambda}\right)\right] \cos 2k_iR_j.
							\label{eq:ACoscoeff}							
\end{eqnarray}
We also set the coefficient vector $\boldsymbol{w}$ as follows:  
\begin{eqnarray}
\boldsymbol{w}= [w_0, w_{1,0},w_{1,1},\dots,w_{j,0},w_{j,1},\dots,w_{M,0},w_{M,1}]^\mathrm{T},   
\end{eqnarray}
where $w_0$ represents an intercept. Although many values have been reported for the mean free path of copper, ranging from $5.2$~\AA \cite{SternPhysRevB1975} to over $10$~\AA \cite{FornasiniPhysRevB2004}, we select $10$~\AA~for the constant value of $\Lambda$ \cite{Akai2018}.
From the coefficients $w_{j,0}$ and $w_{j,1}$, 
the quasi-coordination number $\hat{N}(R_j)$ is obtained as follows: 
$\hat{N}(R_j)\propto \sqrt{\{w_{j,0}\}^2+\{w_{j,1}\}^2}. 
\label{eq:NRj_propto}$

Next, let us consider a conventional basis, the FT basis. 
As denoted below, the FT basis is represented by $\sin2kR_j$ and $\cos2kR_j$, and 
similar to AdvFT, each element of the FT basis function matrix $\boldsymbol{X}^{\mathrm{FT}}$ 
is simply written as follows:
\begin{eqnarray}
X_{i, j, 0}^{\mathrm{FT}}= \sin 2k_iR_j,\label{eq:FTSincoeff} \;
X_{i, j, 1}^{\mathrm{FT}}=\cos 2k_iR_j.\label{eq:FTCoscoeff}
\end{eqnarray}
Fig. \ref{fig:Exp_FT_WT_AdvFT}(b) shows the sine function of the FT basis with fixed $R_j=2$~\AA, which has a maximum magnitude in the magnitude of the FT. 

Finally, we introduce the WindowedFT basis, as shown in Fig. \ref{fig:Exp_FT_WT_AdvFT}(c). 
Similar to a window operation such as a windowed FT, the WindowedFT basis is a wave packet, and it was incorporated in previous works for EXAFS analysis  \cite{funke2005wavelet,timoshenko2009wavelet,timoshenko2014exafs,timoshenko2018neural}. 
In this study, 
similar to the Morlet wavelet \cite{grossmann1990reading,funke2005wavelet},
windowed Fourier (WindowedFT) basis is selected \cite{funke2005wavelet}. 
WindowedFT is obtained by taking a complex sine wave (as in FT) and confining it with a Gaussian envelope. 
Each element of the WindowedFT basis function matrix $\boldsymbol{X}^{\mathrm{WindowedFT}}$ 
is simply written as follows:
\begin{eqnarray}
  X_{i, j, 0}^{\mathrm{WindowedFT}}=
\exp\left[-2\left(k_i-k_{\mathrm{center}}\right)^2\sigma_{\mathrm{DW}}^2\right] \sin 2k_iR_j,
							\label{eq:WTSincoeff}  
\end{eqnarray}
\begin{eqnarray}
  X_{i, j, 1}^{\mathrm{WindowedFT}}=
\exp\left[-2\left(k_i-k_{\mathrm{center}}\right)^2\sigma_{\mathrm{DW}}^2 \right] \cos 2k_iR_j, 
							\label{eq:WTCoscoeff}	  
\end{eqnarray}
where $k_{\mathrm{center}}$ represents 
a central wavenumber of the wave packet with a symmetrical envelope shape and $\sigma_{\mathrm{DW}}$ is the scale value, which is closely related to the Debye-Waller factor. All the basis functions defined by Eqs.(8) and (9) have the center of mass at the same position defined by $k_{\mathrm{center}}$ in Windowed Fourier basis, while in wavelet transform, $k_{\mathrm{center}}$ assume all possible values from $k_0$ to $k_p$ for a complete wavelet basis set and the width of the Gaussian function (defined by $\sigma_{\mathrm{DW}}$ in Eqs. (8) and (9)) scale with frequency $R_j$. Although Eqs. (\ref{eq:ASincoeff}) and (\ref{eq:ACoscoeff}) provide an asymmetric envelope shape for EXAFS oscillations due to the terms of $k^3$ and the Debye-Waller factor, the WindowedFT basis has a symmetric envelope shape, as shown in Eqs. (\ref{eq:WTSincoeff}) and (\ref{eq:WTCoscoeff}). 
This is thought to affect the estimation of the Debye-Waller factor and the relevance of WindowedFT analysis for EXAFS signals. 

\begin{figure*}[t!]
  \centering
  \includegraphics[width=6in]{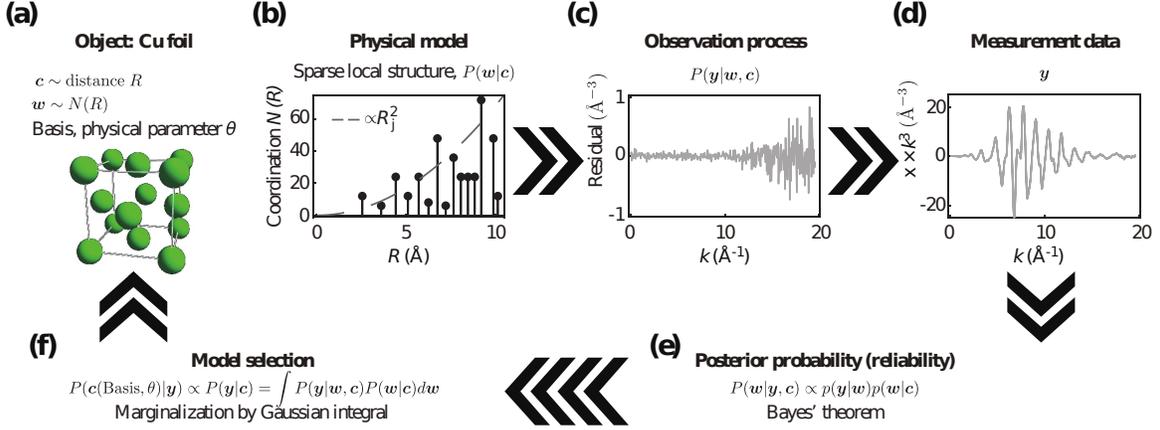}
  \caption{Bayesian analysis for EXAFS data}
  \label{fig:BayesIntro}
\end{figure*}

\subsection{Sparse regression analysis for EXAFS signal}
Let us revisit the purpose of EXAFS analysis and relate it to the machine learning problem, namely, regression analysis. To derive the information about atomic-scale local structures from the EXAFS signals $\boldsymbol{y}$, we solve the regression problem $\boldsymbol{y}$ using the denoted basis $\boldsymbol{X}$, as shown in Eq. (\ref{eq:chik3=Ax}), and derive the coefficient vector $\boldsymbol{w}$. 
As indicated above, the proportional quasi-coordination number is derived from $\hat{N}(R_j)\propto \sqrt{\{w_{j,0}\}^2+\{w_{j,1}\}^2}$. When each coefficient $w_{j,l}$ $(\forall l=\{0,1\})$ is nonzero, an atom exists at distance $R_j$. There is an underlying physical assumption that the atomic arrangement in a substance is actually discrete (i.e., sparse). 

We then formulate the linear regression problem by using an indicator variable that represents a combination of nonzero basis functions, which is often called a sparse linear regression. 
The indicator is defined as a $(2M+1)$-dimensional binary vector $\boldsymbol{c} \in \{0,1\}^{2M+1}$ as follows:
\begin{eqnarray}
\boldsymbol{c} =(c_{0},c_{1,0}, c_{1,1}, \dots, c_{j,0}, c_{j,1}, \dots,  c_{M,0}, c_{M,1}), 
\end{eqnarray}
where we set $c_0=1$ and each variable $c_{j,l}$ takes $0$ or $1$: $c_{j,l}=1$ if the $i$ and $l$th variables belong to the combination and $c_{j,l}=0$ otherwise. 
Using the indicator vector $\boldsymbol{c}$, we can write the sparse linear regression problem as follows: 
\begin{eqnarray}
\boldsymbol{y} = \boldsymbol{X} (\boldsymbol{c} \circ \boldsymbol{w}) + \boldsymbol{\epsilon}, 
\label{eq:ESLR1}
\end{eqnarray}
where the symbol $\circ$ represents the Hadamard product, namely, $(\boldsymbol{c}\circ\boldsymbol{w})_{j,l}=c_{j,l} w_{j,l}$. 

This formulation makes the essence of the problem more explicit, and we can search for the best combination $\boldsymbol{c}$ and its coefficients, $\boldsymbol{w}$, by evaluating the ability of regression for objective variables $\boldsymbol{y}$. 
Moreover, the regression results are affected by the basis $\boldsymbol{X}$, such as the FT, WindowedFT and AdvFT bases, and the physical parameters, such as the Debye-Waller factor. Thus, by comparing and evaluating the results of each basis or the physical parameters, we can select the best basis and physical parameters for the regression of the EXAFS signal, as discussed in the next section.

\section{Method}
\subsection{Bayesian inference for sparse linear regression}
In this paper, we use Bayesian inference for the sparse linear regression problem in EXAFS analysis. 
Fig. \ref{fig:BayesIntro} shows the outline of Bayesian inference in EXAFS analysis. 
There is good prior information on the physical model (Fig. \ref{fig:BayesIntro}(b)) and observation process (Fig. \ref{fig:BayesIntro}(c)), 
as becomes apparent below in Subsection \ref{subsec:Prior}. 
Since the Bayesian approach allows this information to be incorporated into the statistical analysis as a probabilistic model \cite{bishop2006pattern}, 
we can not only determine the reliability of 
the magnitude spectra (Fig. \ref{fig:BayesIntro}(e)) \cite{Krappe2000,Krappe2002}  
but also select the appropriate model (Fig. \ref{fig:BayesIntro}(f)), such as the basis function and physical parameters, from EXAFS signals (Fig. \ref{fig:BayesIntro}(d)).

Let us introduce Bayesian inference in concrete terms. 
 First, we consider the linear regression problem of deriving $\boldsymbol{w}$ on the basis of the given $\boldsymbol{c}$, which represents the distance between the atomic-scale local structures and the target structure. 
In Bayesian inference, this linear regression problem is optimized by maximizing the posterior probability $P(\boldsymbol{w}|\boldsymbol{y},\boldsymbol{c})$ as follows \cite{bishop2006pattern}:
\begin{eqnarray}
\hat{\boldsymbol{w}} = \argmax_{\boldsymbol{w}}P(\boldsymbol{w}|\boldsymbol{y},\boldsymbol{c}). 
\label{eq:MAPDef}
\end{eqnarray}
Based on the Bayes theorem \cite{bishop2006pattern}, 
the posterior probability is derived 
from the product of the prior probability  $P(\boldsymbol{w}|\boldsymbol{c})$ and likelihood $P(\boldsymbol{y}|\boldsymbol{w},\boldsymbol{c})$. 
\begin{eqnarray}
P(\boldsymbol{w}|\boldsymbol{y},\boldsymbol{c}) 
&=& 
\frac{P(\boldsymbol{y}|\boldsymbol{w},\boldsymbol{c})
P(\boldsymbol{w}|\boldsymbol{c})}{P(\boldsymbol{y}|\boldsymbol{c})} \\
&\propto& P(\boldsymbol{y}|\boldsymbol{w},\boldsymbol{c})
P(\boldsymbol{w}|\boldsymbol{c}). 
\label{eq:posterior}
\end{eqnarray}
In the next section, we present clear details about a proper prior probability $P(\boldsymbol{w}|\boldsymbol{c})$ and likelihood $P(\boldsymbol{y}|\boldsymbol{w},\boldsymbol{c})$ for EXAFS analysis. 

Based on the Bayesian inference, we then consider the estimation of $\boldsymbol{c}$ from EXAFS signals, which represents the atomic-scale local structures and is defined by whether each coefficient $w_{j,l} \; (\forall l=\{0,1\})$ is zero. The estimation of $\boldsymbol{c}$ is precisely determined by exhaustively evaluating all combinations of $p$ explanatory variables in terms of a certain information criterion. The total number of all the combinations to be searched is $2^p$, which is called the exhaustive search (ES) method \cite{Igarashi2018}. The ES method has a computational complexity of $O(2^p)$ \cite{Cover1977}. The ES method easily becomes intractable for a large $p$. 
To reduce the computational load, a relaxation approach for the sparse variable selection is used, such as the least absolute shrinkage and selection operator (LASSO) method \cite{Tibshirani1996}, where $L1$-regularization is used to implement the sparseness in the atomic coordination, as discussed below. 

LASSO is formulated as a least squares method with $L1$-regularization as follows:
\begin{eqnarray}
  \hat{\boldsymbol{w}}(\lambda)
= \argmin_{\boldsymbol{w}}
						\frac{1}{2}(\boldsymbol{y}- \boldsymbol{X}\boldsymbol{w})^\mathrm{T}\Sigma^{-1}
						(\boldsymbol{y}- \boldsymbol{X}\boldsymbol{w})
								+ \lambda ||\boldsymbol{w}||_1,
\label{eq:LASSO}
\end{eqnarray}
where $\|\cdot\|_1$, called the ``$L1$-norm'', is defined as $\|\boldsymbol{w}\|_1=\sum_i|w_i|$, and its coefficient $\lambda$ is called a ``regularization parameter'', 
which should be handled very carefully. If $\lambda$ is set to a moderate value, LASSO suppresses some of the coefficients $\boldsymbol{w}$ to zero and leads to an appropriate sparse combination of explanatory variables. 
Otherwise, an excessively sparse combination or a nonsparse combination can be obtained. 

Here, a previous work\cite{ershov1981new} used both ``$L2$-norm'' regularization, defined by $\|\boldsymbol{w}\|_2=\sum_i(w_i)^2$, and total variation, defined by the difference between the adjacent weights namely $\sum_i(w_{i+1}-w_i)^2$. The resulting model weights tend to be smaller in absolute value. The difference between the neighboring weights becomes small, and a smooth solution is obtained. In this case, we can't get sparse solutions like those obtained by L1 regularization. 
According to the notation of the indicator vector $\boldsymbol{c}$, it is convenient that the combination selected by LASSO with $\lambda$ is denoted by $\boldsymbol{c}(\lambda)$: $c_i(\lambda)=1$ if $\hat{w}_i(\lambda)\not=0$ and 
 $c_i(\lambda)=0$ otherwise. 

To select an optimized $\boldsymbol{c}(\lambda)$, which shows an atomic arrangement in a substance, 
we can perform model selection within the framework of Bayesian inference \cite{bishop2006pattern}. 
Here, the model is represented by $\boldsymbol{c}(\lambda)$, and we calculate a ranking based on the 
Bayesian free energy (BFE) of each model $-\log P(\boldsymbol{y}|\boldsymbol{c})$, 
which is the negative logarithmic posterior probability marginalized 
by $\boldsymbol{w}$, as presented in Subsection \ref{subsec:BFEcalc}. The BFE minimization is identical to the posterior probability maximization. 

Based on the BFE, we can optimize the parameter of $L1$-regularization $\lambda$ and obtain a sparse combination of bases for EXAFS signals, which is called Bayesian LARS-OLS \cite{Efron2004,Igarashi2018,Mototake2018}. 
We can obtain an optimized $\lambda$, which leads to an appropriate sparse solution because BFE is minimized. 
If $\lambda$ is set to a large value, LASSO suppresses almost all the coefficients $\boldsymbol{w}$ to zero, 
which indicates that the extracted bases are close to zero and 
leads to a large BFE because the model cannot fit the EXAFS signals. 
However, if $\lambda$ is set to a small value, 
the number of extracted bases is large, and the excessively complex function could be fit by this model. 
Due to the complexity, this leads to reducing the probability of a moderate model for fitting the EXAFS signals and to increasing the BFE of the model. 
Thus, this BFE can extract simple (sparse) combinations of 
bases for fitting EXAFS signals \cite{bishop2006pattern}. 
It is known that BFE can select a model sparser than those extracted by cross-validation error \cite{Igarashi2018,Mototake2018}. 
Moreover, the selected sparse model in EXAFS analysis 
is definitely dependent on the basis function, such as FT, WindowedFT and AdvFT, and a physical parameter. 
Thus, we can expand the model selection to investigate the basis selection and optimize the physical parameter, namely, the Debye-Waller factor in this paper. 

\subsection{Prior knowledge for Bayesian inference}
\label{subsec:Prior}
To calculate the posterior probability $P(\boldsymbol{w}|\boldsymbol{y},\boldsymbol{c})$ and BFE $-\log P(\boldsymbol{y}|\boldsymbol{c})$ based on a physical model and observation process, we propose a specific prior probability $P(\boldsymbol{w}|\boldsymbol{c})$ and likelihood $P(\boldsymbol{y}|\boldsymbol{w},\boldsymbol{c})$. 

\subsubsection{General characteristics in coordination number }
\begin{figure}[t!]
  \centering
  \includegraphics[width=4in]{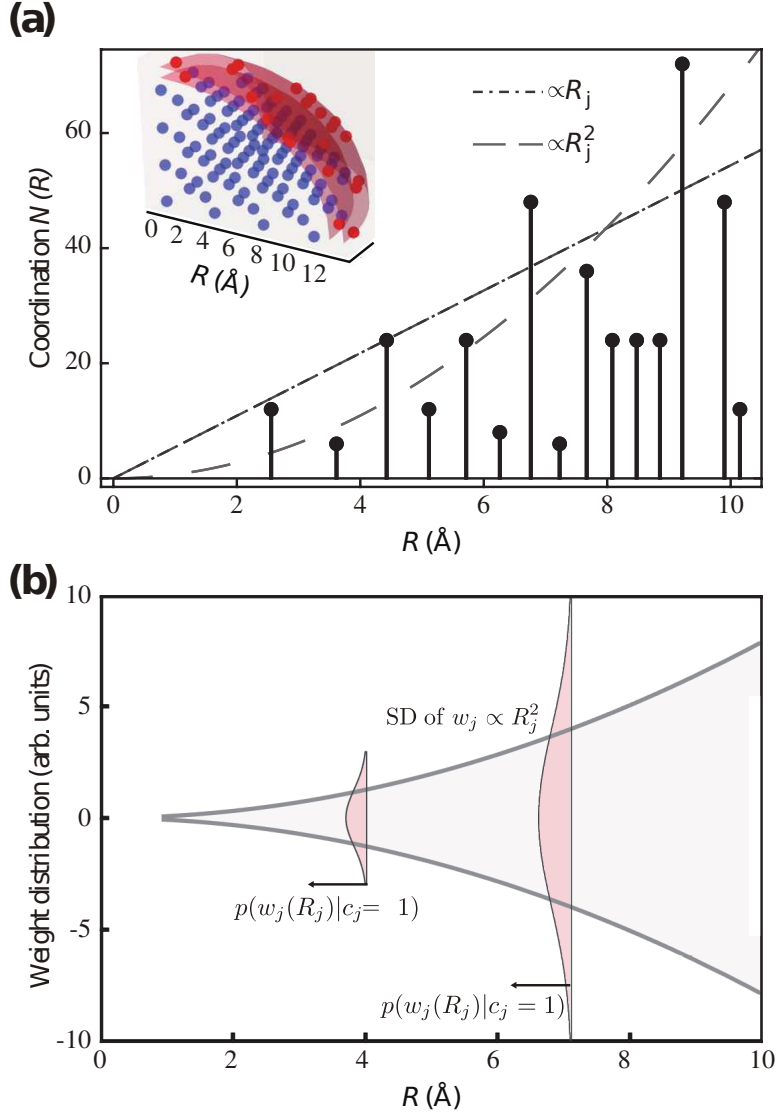}
  \caption{(a) Coordination number of fcc copper. 
  (b) Weight distribution depending on the distance between neighboring atoms. 
  The radial distribution function of an fcc copper is derived with a lattice constant of $a=3.61$~\AA~at $T=300~K$ \cite{hermann2017crystallography}. 
  The nearest-neighbor atoms are located at the symmetrically equivalent relative positions to
  $\left[\frac{1}{2}a, \frac{1}{2}a, 0\right]$ and their atomic distance is $2.55$ \AA $(=a/\sqrt{2})$ in copper. 
  }
  \label{fig:weight_prior}
\end{figure}

Since the number of neighboring atoms increases with increasing distance from a target atom in solid and liquid materials, 
copper's coordination number $N(R)$ increases as the distance $R$ increases,
as shown in Fig. \ref{fig:weight_prior}(a). 
Moreover, since the coordination number among the thin sphere shell in $R_{j}\sim{}R_{j}+\Delta{}R$ is assumed to be proportional to the square of the distance $R_{j}^2$ in the three-dimensional isotropic continuum, as shown in the inset of Fig. \ref{fig:weight_prior}(a), 
we expect that the $N(R)$ of a highly symmetric crystal structure is proportional to $R_j^k \; (k= 1 \sim 2)$ at a closer distance $R_j$ and converges to the square of the distance $R_{j}^2$ at a further distance. 

Here, to involve the versatile nature in 
coordination number as a prior probability of coefficient $P(\boldsymbol{w}|\boldsymbol{c})$, 
we introduce that in the case of 
$c_{j,l}=1$, 
$P(w_{j,l}|c_{j,l}=1)$ is a Gaussian distribution, 
where the mean and variance are $0$ and $z_w(R_j)$, respectively, and assume that $z_{\omega}(R_{j})$ is proportional to $(R_{j}^2)^2=R_{j}^4$, as shown in Fig. \ref{fig:weight_prior}(b). This feature of coordination number is mathematically formulated as follows: 
\begin{eqnarray}
P(w_{j,l}|c_{j,l}=1)&=&\sqrt{\frac{z_w(R_j)}{2\pi}}\exp\left(-
\frac{1}{2}z_w(R_j) w_{j,l}^2\right),  \nonumber \\
P(w_{j,l}|c_{j,l}=0) &=& \delta(w_{j,l}),   \nonumber
\end{eqnarray}
where we set 
\begin{eqnarray}
\frac{1}{z_w(R_j)}= \frac{(R_j^2)^2}{z^0_w}= \frac{R_j^4}{z^0_w}
\end{eqnarray}
and
$P(w_0) =\sqrt{\frac{z_w^0}{2\pi}}\exp\left(-z_w^0 w_0^2\right)$. 
The variance $z^0_{w}$ is optimized using the empirical Bayes method, as described in the Appendix 1.

\begin{figure}[t!]
  \centering
  \includegraphics[width=4in]{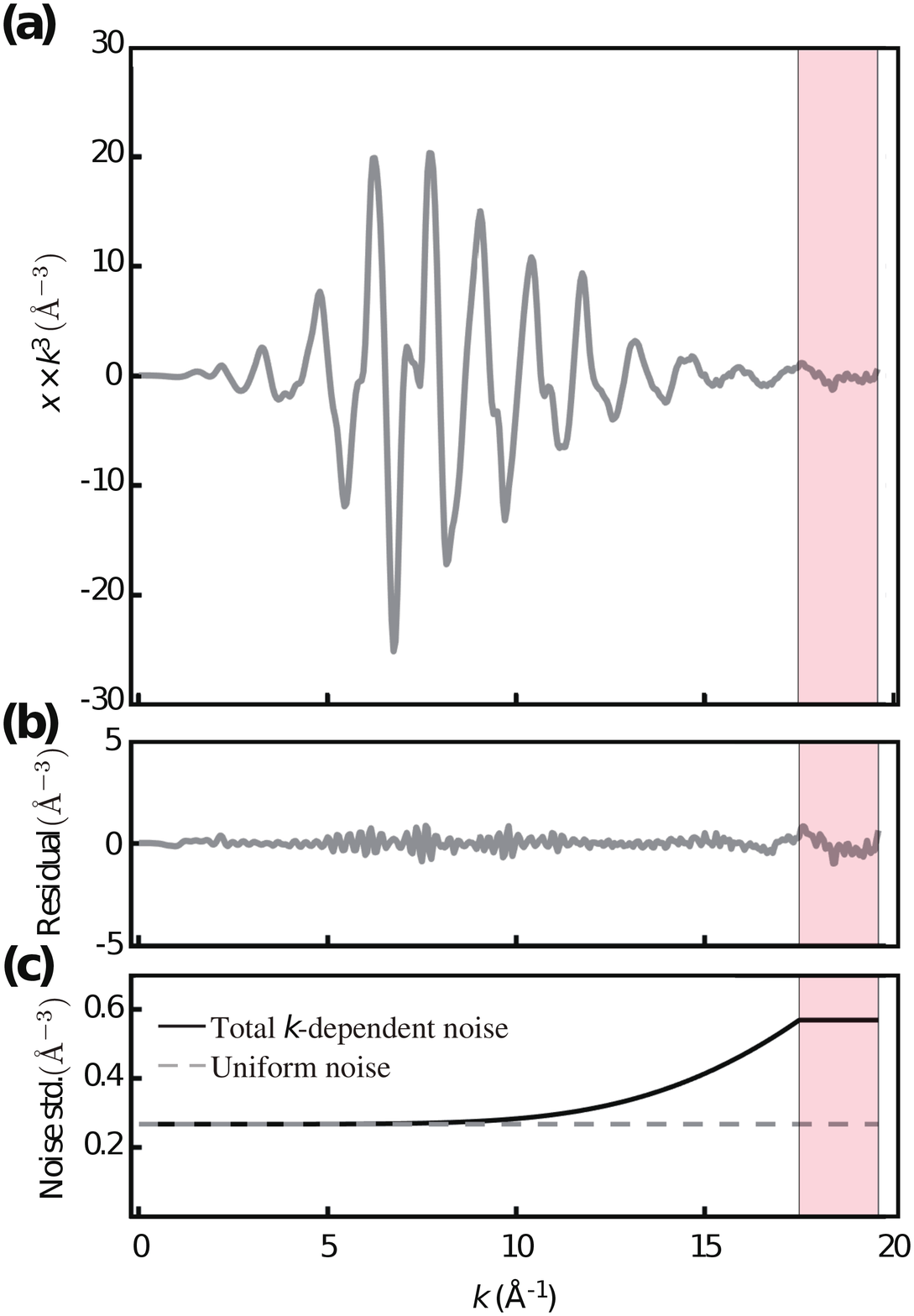}
  \caption{(a) Measured background-subtracted and $k^3$-weighted EXAFS data for copper foil measured at $T=300~K$. (b) Residual of the fit (measured data fit). (c) Noise prior based on the measured data with multiplication of the $k^3$ term. }
  \label{fig:prior}
\end{figure}

\subsubsection{Prior knowledge of noise involved in EXAFS signals}
In this section, we provide a specific likelihood $P(\boldsymbol{y}|\boldsymbol{w},\boldsymbol{c})$ involving observation processes. 
Since an EXAFS signal is often shown by $\chi(k)$ weighted 
with $k^2$ or $k^3$ 
to amplify the oscillations at high-$k$, 
noise should be incorporated with the same weight of $k^2$ or $k^3$ because the noise is considered to be magnified at high $k$.

Assuming that the EXAFS signals at $k =17.5\sim 19.6$~\AA
$^{-1}$~are mainly characterized by measurement noise, 
we calculate the standard variation within this interval and obtain the value $\sigma_\mathrm{max} =0.57$~\AA
$^{-3}$. 
We also assume that the standard variation is $k^3$ weighted. 
Let us consider the effects of $\sigma_\mathrm{max}$ on the obtained basis selection. If $\sigma_\mathrm{max}$ decreases, we treat the EXAFS signals at high $k$ as reliable data and more closely fit the EXAFS signals using many more bases. On the other hand, if $\sigma_\mathrm{max}$ increases, we assume the data reliability at high $k$ is low and conclude that the obtained model does not fit the EXAFS signals at high $k$. 

Additionally, we set another constant noise with the variance $\frac{1}{z}$ that represents systematic noise resulting from the single-scattering approximation for EXAFS signals, and 
this noise is mathematically formulated as follows: 
\begin{eqnarray}
\sigma^2(k_i) = \left(\sigma^2_\mathrm{max} -\frac{1}{z}\right)
\left(\frac{k_i}{k_{\mathrm{max}}}\right)^{3\times2}, 
\label{eq:sigma_k}
\end{eqnarray}
where we set $k_{\mathrm{max}} =17.5$~\AA. 
Using the additivity of variance in the Gaussian distribution and assuming 
that the nondiagonal element $\Sigma_{kl}(k\neq l)= 0$, 
the total variance of signals is represented by 
\begin{eqnarray}
\Sigma_{kk} = \frac{1}{z} + \sigma(k)^2 = \frac{1+ \sigma(k)^2 z}{z}, 
\end{eqnarray}
which is shown in Fig. \ref{fig:prior}(c). 
 Similar to an optimization of the variance $z^0_{w}$, 
the variance $z$ is optimized using the empirical Bayes method, as described in Appendix 1. 

We then set the likelihood $P(\boldsymbol{y}|\boldsymbol{w},\boldsymbol{c})$ 
as a multidimensional Gaussian distribution, where the mean and variance matrix are $0$ and $\Sigma$, respectively, as follows: 
\begin{eqnarray}
P(\boldsymbol{y}| \boldsymbol{w},\boldsymbol{c}) 
=
\frac{1}{\mathrm{det}(2\pi\Sigma)^{1/2}}\\
\times
\exp
\left(-\frac{1}{2}
\left[ \boldsymbol{y}  - \boldsymbol{X} (\boldsymbol{c} \circ \boldsymbol{w})\right]^{\mathrm{T}}\Sigma^{-1}
\left[ \boldsymbol{y}  - \boldsymbol{X} (\boldsymbol{c} \circ \boldsymbol{w})\right]\right).
\label{eq:likelihood}
\end{eqnarray}
The exponential part in Eq. (\ref{eq:likelihood}) shows 
the weighted residual sum of squares. 
Since the weighted residual sum is affected by 
$\Sigma_{kk}$ in the denominator, the residual at high $k$ has an insignificant effect on the weighted residual sum and the likelihood, 
which allows the regression model to be disregarded at high $k$.

\subsection{Radial distribution function based on Bayesian inference}
Using the prior probability $P(\boldsymbol{w}|\boldsymbol{c})$ and likelihood $P(\boldsymbol{y}|\boldsymbol{w},\boldsymbol{c})$, 
we can calculate 
the posterior probability based on Eq. (\ref{eq:posterior}) as follows: 
\begin{eqnarray}
P(\boldsymbol{y}| \boldsymbol{w}, \boldsymbol{c})P(\boldsymbol{w}|\boldsymbol{c}) 
=
\\
\frac{1}{\mathrm{det}(2\pi\Sigma)^{1/2}}\prod_{\forall{j}|c_j=1}^C\sqrt{\frac{z_w(R_j)}{2\pi}}
\exp\Bigg(\\
-\frac{1}{2} 
\Big[ 
(\boldsymbol{c} \circ\boldsymbol{w} - \boldsymbol{\mu})^{\mathrm{T}} \boldsymbol{\Lambda}^{-1}
(\boldsymbol{c} \circ \boldsymbol{w} - \boldsymbol{\mu}) 
+ \boldsymbol{y}^{\mathrm{T}}\Sigma^{-1}\boldsymbol{y}
- \boldsymbol{\mu}^{\mathrm{T}} \boldsymbol{\Lambda}^{-1} \boldsymbol{\mu}
\Big]\Bigg),
\label{eq:Pw_wcPw_c6}
\end{eqnarray}
where $C\equiv|\boldsymbol{c}|$, and
we set $\boldsymbol{\Lambda}^{-1}=\boldsymbol{X}^{\mathrm{T}}\Sigma^{-1}\boldsymbol{X} + \boldsymbol{Z_w}$, $\boldsymbol{Z_w}(j,j)=z_w (R_j)$ $(\forall{j}|c_j=1)$ and $\boldsymbol{\mu}=\boldsymbol{\Lambda} \boldsymbol{X}^{\mathrm{T}} \Sigma^{-1}\boldsymbol{y}$. 
We can then obtain the solution of Eq. (\ref{eq:MAPDef}), which has the maximum a posteriori (MAP), which is an estimate of the coefficient $\boldsymbol{w}$ based on Bayesian inference. The magnitude of estimated $\boldsymbol{w}$ is related to the coordination number. Since the Bayesian inference estimates the probability distribution of $\boldsymbol{w}$ as the posterior probability, we can simultaneously obtain reliability inference for $\boldsymbol{w}$ as the diagonal elements of $\boldsymbol{\Lambda}$, which is also related to the reliability inference of the coordination number. 

Although previous Bayesian approaches used FEFF analysis \cite{Newville2014} to obtain reliable prior knowledge about the measurement object \cite{Krappe2000,Krappe2002}, 
the Bayesian inference framework provides flexibility in the choice of prior knowledge. 
 Specifically, Bayesian inference can combine the above physical prior knowledge and observation process and can solve difficult problems such as atomic-scale local structures, optimization of physical parameters and basis selection without FEFF analysis. 

Let us consider EXAFS data analysis (EDA), which also uses FEFF analysis, as well as the previous study \cite{Krappe2000,Krappe2002} and a study that performs all the steps of the EXAFS analysis \cite{kuzmin2000dehydration}. 
In our formulation, this problem can be treated as a linear regression problem, as shown in Eq. (\ref{eq:ESLR1}), which leads to an analytical solution and fast calculation using Bayesian inference, while in EDA, EXAFS parameters are optimized by a nonlinear least-squares fitting code, which leads to initial value dependence due to local minima and solution instability. 

\subsection{Bayesian free energy for model selection}
\label{subsec:BFEcalc}
To select an optimized $\boldsymbol{c}$, which shows an atomic arrangement in a substance, 
and to extract an appropriate basis function and physical parameter, 
we calculate a ranking based on the BFE of each model $-\log P(\boldsymbol{y}|\boldsymbol{c})$ 
because we assume that $P(\boldsymbol{c})$ is constant, 
and we can calculate $P(\boldsymbol{c}|\boldsymbol{y})$ 
by using Bayes' theorem \cite{bishop2006pattern} as follows: 
\begin{eqnarray}
P(\boldsymbol{c}|\boldsymbol{y}) =
\frac{P(\boldsymbol{y}|\boldsymbol{c})P(\boldsymbol{c})}
{P(\boldsymbol{y})}
\propto P(\boldsymbol{y}|\boldsymbol{c}). 
\end{eqnarray} 
The BFE is the negative logarithmic posterior probability calculated by marginalizing Eq. (\ref{eq:Pw_wcPw_c6})
over $\boldsymbol{w}$ and can be analytically calculated since we provide a specific prior probability $P(\boldsymbol{w}|\boldsymbol{c})$ and likelihood $P(\boldsymbol{y}|\boldsymbol{w},\boldsymbol{c})$ 
as Gaussian distributions. 
 Specifically, the posterior probability is proportional to a marginalized likelihood function defined by
$P(\boldsymbol{y}| \boldsymbol{c})
= \int P(\boldsymbol{y}| \boldsymbol{w},\boldsymbol{c})
P(\boldsymbol{w}|\boldsymbol{c}) d\boldsymbol{w}.
\label{eq:P_marginalization}$
After a straightforward calculation, the resulting formula of the BFE is given as 
\begin{eqnarray}
    \mathrm{BFE}(\boldsymbol{c}) 
    \equiv -\log P(\boldsymbol{y}|\boldsymbol{c}) 
    \label{eq:BFEdif}
    \\
    =\frac{1}{2} [ \boldsymbol{y}^{\mathrm{T}}\Sigma^{-1}\boldsymbol{y}
                        - \boldsymbol{\mu}^{\mathrm{T}} \boldsymbol{\Lambda}^{-1} \boldsymbol{\mu}]
                        +\frac{1}{2}\log[\mathrm{det}(2\pi\Sigma)]
    \\
    -\frac{1}{2}\log\prod_{\forall{j}|c_j=1}^C z_w(R_j)
    + \frac{1}{2}\log\det{\Lambda^{-1}}.
\end{eqnarray}
In the next section, based on the BFE, we optimize $\lambda$ and the physical parameters and select an appropriate basis function for EXAFS analysis from FT, WindowedFT and AdvFT to obtain a sparse combination of bases.

\section{Results and discussion}
\label{sec:Results}

\begin{figure*}[t!]
  \centering
  \includegraphics[width=6in]{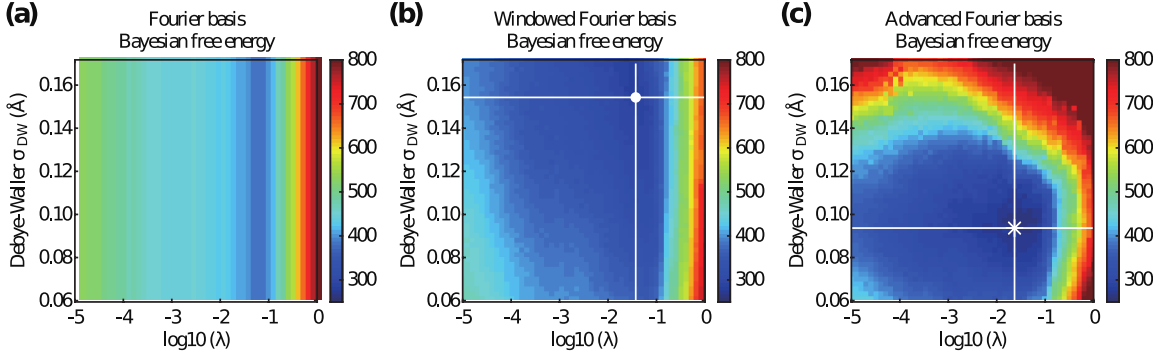}
  \caption{
BFE ($\lambda$, $\sigma_{\mathrm{DW}}$) heatmap presented with a logarithmic abscissa scale of $\lambda$, 
  where the Fourier basis (a), windowed FT 
  (b), and advanced Fourier basis (c) are used in each heatmap. 
  The circle and asterisk points in Figs. (b) and (c) show the minimum BFE using WindowedFT and AdvFT bases, respectively. 
  Figs. \ref{fig:HyperParamEst-DWsigmavslambdaCVEandFEmap-slices}(a) and (b) present the cross-sections, which 
  are sliced at the minimal BFE points along the white lines in Figs. (b) and (c).
 }
  \label{fig:HyperParamEst-DWsigmavslambda-FEmap}
\end{figure*}

\begin{figure*}[t!]
  \centering
  \includegraphics[width=6in]{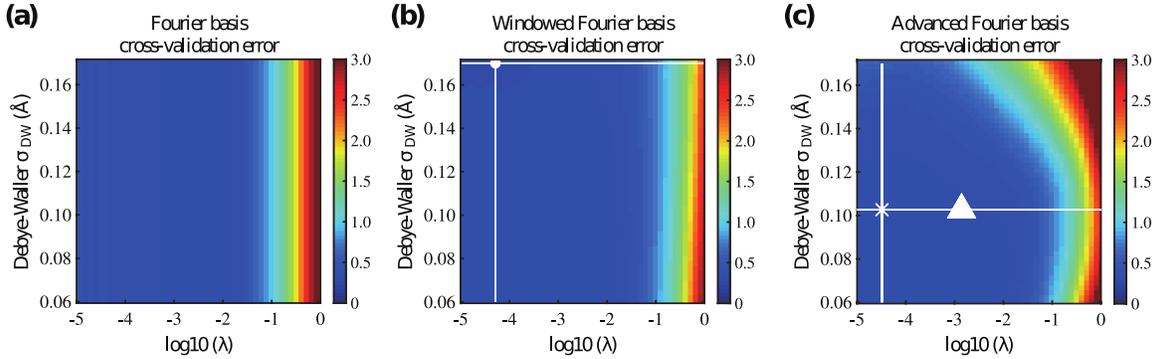}
  \caption{
CVE ($\lambda$, $\sigma_{\mathrm{DW}}$) heatmap presented with a logarithmic abscissa scale of $\lambda$, 
  where the Fourier basis (a), windowed FT (b), and advanced Fourier basis (c) are used in each heatmap. 
  The circle and asterisk points in Figs. (b) and (c) show the minimum CVE using WindowedFT and AdvFT bases, respectively. 
  The triangle point in Fig. (c) represents the point optimized by the 1SE rule \cite{Murphy2012}. 
  Figs. \ref{fig:HyperParamEst-DWsigmavslambdaCVEandFEmap-slices}(c) and (d) present the cross-sections that 
  are sliced at the minimal CVE points along the white lines in Figs. (b) and (c). 
 }
 \label{fig:HyperParamEst-DWsigmavslambda-CVEmap}
\end{figure*}

\begin{figure*}
  \centering
  \includegraphics[width=6in]{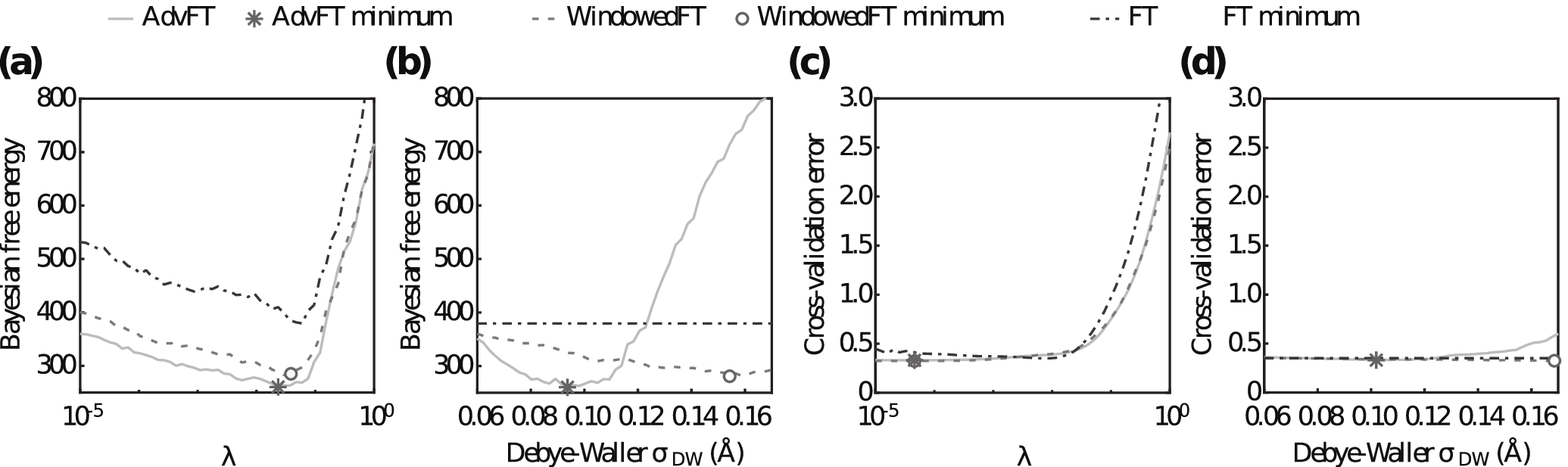}
  \caption{
Comparison of the results of the Fourier basis (dash-dot-dash line), windowed FT (dash line), and advanced Fourier basis (solid line). These results are the slice taken in Figs.~\ref{fig:HyperParamEst-DWsigmavslambda-FEmap} and \ref{fig:HyperParamEst-DWsigmavslambda-CVEmap}. 
In each figure, the minimized BFE and CVE with AdvFT, WindowedFT and FT are represented by 
asterisk, circle and diamond marks, respectively. 
(a), (b): Comparison of the results based on BFE as a function of $\lambda$, $\sigma_{\mathrm{DW}}$, respectively. (c), (d): Comparison of the results based on CVE as a function of $\lambda$, $\sigma_{\mathrm{DW}}$, respectively.}
  \label{fig:HyperParamEst-DWsigmavslambdaCVEandFEmap-slices}
\end{figure*}

\subsection{Physical parameter optimization and basis selection based on Bayesian inference}
\label{subsec:BasisSelection}
Let us consider the optimization of the $L1$-regularization parameter $\lambda$ and the Debye-Waller factor $\sigma_{\mathrm{DW}}$ in each basis based on BFE. 
Fig. \ref{fig:HyperParamEst-DWsigmavslambda-FEmap} shows a two-dimensional heat map for the optimization of $\lambda$ and $\sigma_{\mathrm{DW}}$ using three bases: FT, WindowedFT and AdvFT. 
The FT basis does not include $\sigma_{\mathrm{DW}}$, which results in a constant BFE and CVE regardless of this parameter, as shown in Fig \ref{fig:HyperParamEst-DWsigmavslambda-FEmap}(a), and thus, this basis cannot be used for the optimization of $\sigma_{\mathrm{DW}}$. 
However, WindowedFT and AdvFT are both dependent on $\sigma_{\mathrm{DW}}$, as shown in Eqs. (\ref{eq:ASincoeff}), (\ref{eq:ACoscoeff}), (\ref{eq:WTSincoeff}) and (\ref{eq:WTCoscoeff}). We then conduct a two-dimensional grid search for the optimization of the $L1$-regularization parameter $\lambda$ and the Debye-Waller factor $\sigma_{\mathrm{DW}}$, as shown in Figs. \ref{fig:HyperParamEst-DWsigmavslambda-FEmap}(b) and (c). 
In the WindowedFT basis, there is another parameter that is a central wavenumber $k_{\mathrm{center}}$, and we thus conduct a three-dimensional grid search for the optimization of $\lambda$, $\sigma_{\mathrm{DW}}$ and $k_{\mathrm{center}}$. 

With the AdvFT basis, the optimized Debye-Waller factor $\hat{\sigma}_{\mathrm{DW}}$ based on BFE is $0.094$~\AA, as denoted by the asterisk point
in Fig.~\ref{fig:HyperParamEst-DWsigmavslambda-FEmap}(c). 
This coincides well with the Debye-Waller factor $\hat{\sigma}_{\mathrm{DW}}$ of the first shell derived by FEFF \cite{Newville2014} in terms of accuracy. 
By using the FEFF software, the Debye-Waller factor is estimated to be $0.092\pm0.011$~\AA~for the nearest-neighbor atoms, in which the assumption of a face-centered-cubic (fcc) copper structure is required. 
Since we use a single-scattering approximation for the analysis, the value is close to the Debye-Waller factor $\hat{\sigma}_{\mathrm{DW}}$ of the first shell because the main contribution comes from the nearest neighbors. 
This result indicates that our method can extract the decoherence effect on the dominant component of the EXAFS oscillations included in the measured EXAFS data. 
However, with the WindowedFT basis, there are two 
optimized parameters: 
$\hat{\sigma}_{\mathrm{DW}}$ and $k_{\mathrm{center}}$. The latter parameter represents the central wavenumber of the wave packet with a symmetrical envelope shape and is optimized as $k_{\mathrm{center}}=8$~\AA$^{-1}$ 
(data not shown). 
These results indicate that the optimized center of the WindowedFT basis is the center of the dominant component of the EXAFS oscillations, as shown in Fig.~\ref{fig:Exp_FT_WT_AdvFT}(c). 
Additionally, the optimized Debye-Waller factor $\hat{\sigma}_{\mathrm{DW}}$ based on BFE is $0.16$~\AA,~as denoted by the circle point 
in Fig. \ref{fig:HyperParamEst-DWsigmavslambda-FEmap}(b), which is an outlier of the FEFF result. Since there is a discrepancy between the WindowedFT basis and EXAFS signals in terms of symmetric or asymmetric properties, this basis appears to lead to mismatched results. 
Thus, the estimation of the Debye-Waller factor cannot be realized by WindowedFT or FT but can be realized by AdvFT within the framework of Bayesian inference. 

By comparing the minimum BFE of each basis in Figs. \ref{fig:HyperParamEst-DWsigmavslambda-FEmap}(a), (b) and (c), we can select the physical parameters and an appropriate basis for the regression of EXAFS signals. 
Figs. \ref{fig:HyperParamEst-DWsigmavslambdaCVEandFEmap-slices}(a) and (b), which present the cross-sections of Fig. \ref{fig:HyperParamEst-DWsigmavslambda-FEmap}, show the dependence of the BFE of each basis on the common logarithmic abscissa of the $L1$-regularization parameter $\lambda$ and the Debye-Waller factor in the common ordinate scale. 
In Figs. \ref{fig:HyperParamEst-DWsigmavslambdaCVEandFEmap-slices}(a) and (b), the BFEs minimized with AdvFT, WindowedFT and FT are represented by 
asterisk, circle and diamond marks, respectively. 
We then find that the BFEs with AdvFT, WindowedFT and FT are $260$, $280$, and $379$, respectively, 
and the corresponding posterior probabilities among all the models are $69\%$, $8.5\times10^{-8}\%$ and $1.3\times10^{-50}\%$. 
Since the BFE is a negatively exponential component, as shown in Eq. (\ref{eq:BFEdif}), 
a slight difference in BFE values greatly affects the posterior probability of each model, as referred to above. 
In this way, the Bayesian approach can optimize the physical parameters and basis selection for each EXAFS signal in a quantitative way. Thus, using BFE, AdvFT is selected as an appropriate basis function among FT, WindowedFT and AdvFT for the EXAFS signals of copper. The behavior of the EXAFS oscillation changes dramatically depending on the crystal structure. Therefore, further investigation is needed to generalize the applicability scope of AdvFT. 
These results are summarized by Table \ref{BFE_Basis_summary}. 
\begin{table*}[h!]
\begin{center}
  \caption{Results by Bayesian inference for copper}
   \begin{tabular}{| l | c | c | c | c |} 
   \hline
  Basis & FT & WindowedFT & AdvFT & Remarks\\ 
  \hline \hline
  BFE min.        & $379$ & $280$ & $260$ & \\ 
  Posterior prob. & $1.3\times10^{-50}\%$ & $8.5\times10^{-8}\%$ & $69\%$ & \\ 
  \# of non-zero elements  &  $96$ &  $75$ &  $82$  & \\
  $\sigma_{\mathrm{DW}}$ & - & $0.16$~\AA~ & $0.094$~\AA~ & FEFF: $0.0992\pm 0.011$~\AA~\\
   \hline
  \end{tabular}
  \label{BFE_Basis_summary}
  \end{center}
\end{table*}

To demonstrate the advantage and robustness of the Bayesian approach for optimization and basis selection, we compare the results obtained on the criterion (basis) of the cross-validation error (CVE) \cite{Akai2018} 
in Fig. \ref{fig:HyperParamEst-DWsigmavslambda-CVEmap}. 
In Appendix 2
, we evaluate the approximate prediction performance of the CVE for EXAFS signals. 
To clearly compare the two criteria for the optimization, Fig. \ref{fig:HyperParamEst-DWsigmavslambdaCVEandFEmap-slices} presents the cross-sections of Figs. \ref{fig:HyperParamEst-DWsigmavslambda-FEmap} and \ref{fig:HyperParamEst-DWsigmavslambda-CVEmap}. This figure shows the dependence of the BFE and CVE on the common scales in both the ordinate and  abscissa, where the cross-sections are sliced at the minimal BFE points along the white lines in Figs. \ref{fig:HyperParamEst-DWsigmavslambda-FEmap}(a), (b) and (c).

Although the BFE varies drastically with both $\lambda$ and the Debye-Waller factor, as shown in Figs. \ref{fig:HyperParamEst-DWsigmavslambdaCVEandFEmap-slices}(a) and (b),
and there are clear minimum points as explained above, 
the CVE only shows gradual variations in the Debye-Waller factor and minimum points, 
as shown in Figs. \ref{fig:HyperParamEst-DWsigmavslambdaCVEandFEmap-slices}(c) and (d). 
Because the standard deviation of the minimum CVE is 
$2.4$~$\times$~$10^{-2}$ at $\lambda = 3.2$~$\times$~$10^{-5}$ 
and $\sigma_{\mathrm{DW}}=0.103$~\AA, 
the Debye-Waller factor whose CVE is within the one standard error above the minimal CVE covers a wide range from $0.07$ to $0.123$~\AA. 
This leads to uncertainty in parameter optimization depending on the measured data and random partitioning for the CVE \cite{Akai2018}. 
Moreover, the minimum CVE with AdvFT
differs minimally from those with FT and WindowedFT when comparing the standard deviation of the CVE, as shown in Figs. \ref{fig:HyperParamEst-DWsigmavslambdaCVEandFEmap-slices}(c) and (d). 
Thus, the BFE has more advantages for optimization and basis selection than the CVE \cite{Akai2018}. 

\begin{figure*}
  \includegraphics[width=6in]{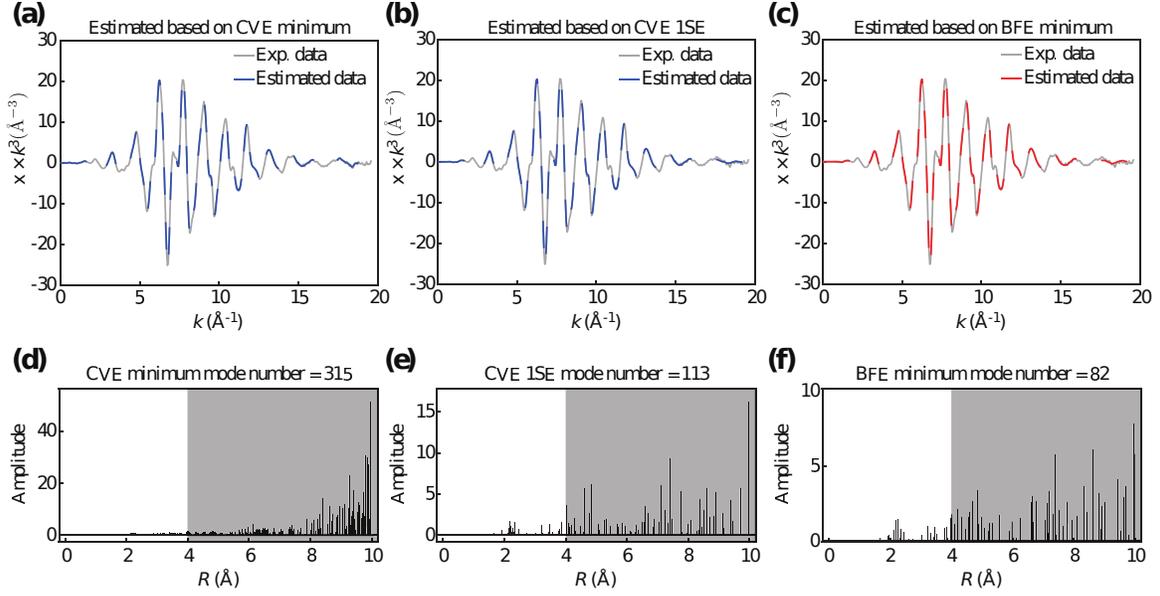}
  \caption{
 (a), (b), (c): Measured background-subtracted and $k^3$-weighted EXAFS data for copper (solid line) and fitted data (dashed line) using advanced Fourier basis (AdvFT). 
  (d), (e), (f): Magnitude spectra obtained by AdvFT. 
  In each figure, the $L1$-regularization parameter $\lambda$ and the Debye-Waller factor are optimized by CVE minimization (a) and (d) ($\sigma_{\mathrm{DW}}=0.103$~\AA, $\lambda_{\mathrm{CVEmin}}=3.2\times 10^{-5}$), the CVE 1SE rule (b) and (e) ($\sigma_{\mathrm{DW}}=0.103$~\AA, $\lambda_{\mathrm{CVE1SE}}=2.1\times 10^{-3}$), and BFE minimization (c) and (f) ($\sigma_{\mathrm{DW}}=0.094$~\AA~ and $\lambda_{\mathrm{BFE}}=2.3\times 10^{-2}$), respectively. 
  }
  \label{fig:BFEandCVE1SE_Est_mode}
\end{figure*}

The proposed algorithm was raw coded in MATLAB. It should be noted that all the experiments were conducted on two 2.2 GHz Intel(R) Xeon(R) Platinum 8276 CPUs with 28 cores each, namely, 56 cores were used in the calculation. 
Because both $\lambda$ and the Debye-Waller factor are optimized in parallel, it takes 20 minutes to optimize the two parameters by exhaustive research, as shown in Fig. \ref{fig:HyperParamEst-DWsigmavslambda-FEmap} (c). The computational time for optimizing these parameters can be greatly reduced by Bayesian optimization \cite{snoek2012practical}. Moreover, the computational cost of the proposed method using BFE is greatly reduced compared to that by CVE, as shown in Fig. \ref{fig:HyperParamEst-DWsigmavslambda-CVEmap} (c), which requires repetitive calculations, as explained 
in Appendix 2 
\cite{Mototake2018}. 
Finally, if you have optimized physical parameters such as the Debye-Waller factor, only 1.5 minutes is required for quantitative analysis of a single experimental Cu EXAFS spectra, as denoted in Figs. \ref{fig:BFEandCVE1SE_Est_mode}(c) and (f). 

 \begin{figure}
  \centering
  \includegraphics[width=4in]{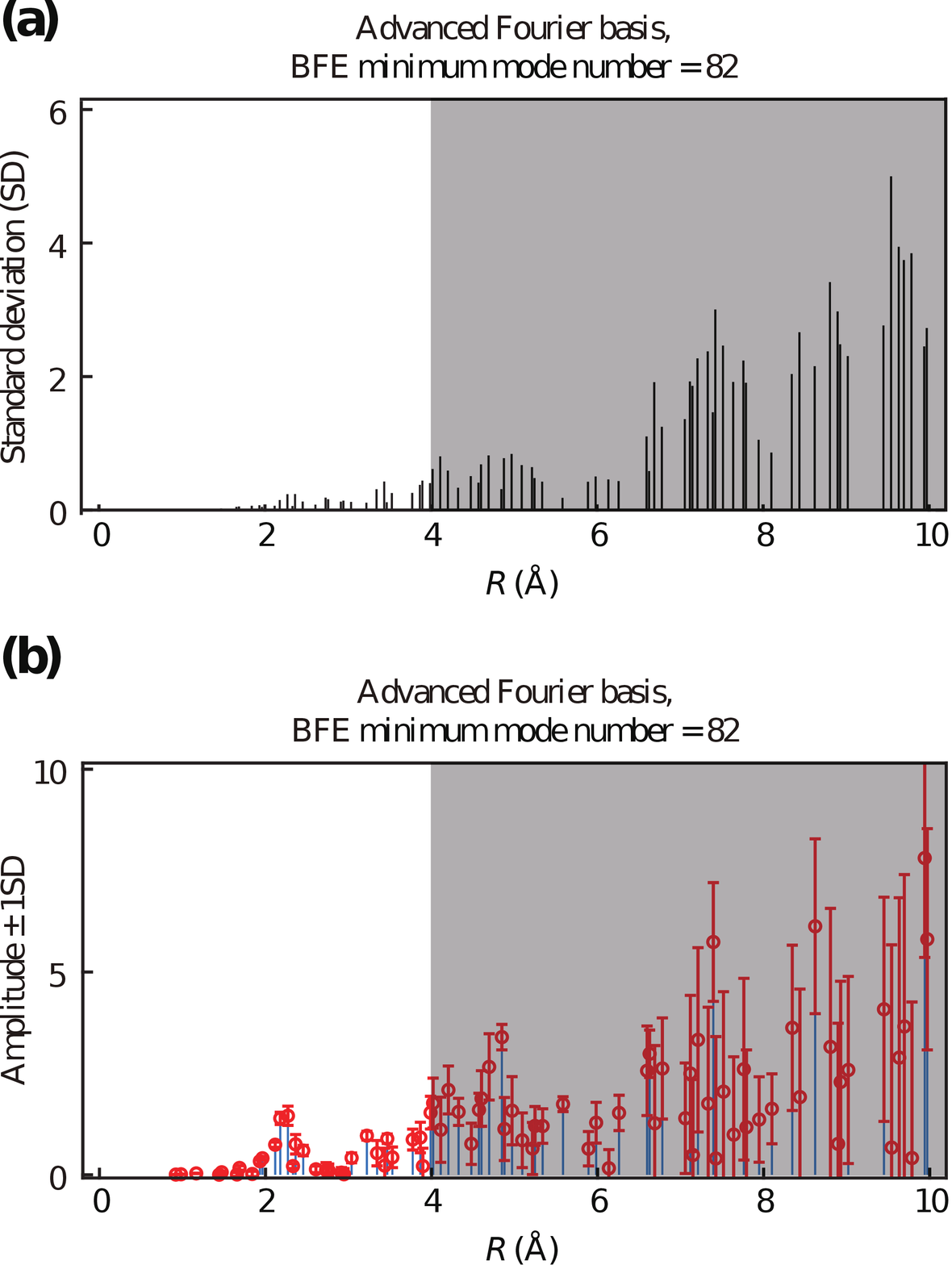}
  \caption{
Reliability of our analysis according to the errors in the coordination number based on Bayesian inference. 
(a) Estimated standard deviation of the amplitude, which is closely correlated to the coordination number.
(b) Magnitude spectrum with the standard deviation. 
    }
  \label{fig:VarEstConfidence}
\end{figure}

\subsection{Magnitude spectra obtained by AdvFT basis based on Bayesian inference}
\label{subsec:Fit}
Using the selected AdvFT basis and the two optimized parameters in the previous subsection, 
we can then obtain the regression results and magnitude spectra obtained by AdvFT basis based on Bayesian inference. 
Figs. \ref{fig:BFEandCVE1SE_Est_mode}(c) and (f) are the respective results that are obtained from the MAP solution with the maximum posterior probability.
As clearly shown in the sparse spectrum in Fig. \ref{fig:weight_prior}(a), the vertical lines representing $N(R)$ for the fcc structure of copper only appear at sparse atomic distance $R$, and the number of basis functions used in the sparse solution is $82$, as shown in Fig. \ref{fig:BFEandCVE1SE_Est_mode}(f), which is only approximately one-fifth of the total number of $R$. 
It is clearly demonstrated that other $R$ components 
with zero intensities are unnecessary for reproducing EXAFS data, 
as shown in Fig. \ref{fig:BFEandCVE1SE_Est_mode}(c). 
The peak positions obtained from Bayesian inference are shifted with respect to the real value of Cu, similar to the previous work \cite{Akai2018}, which discussed concerns about the peak shift in detail. 
Moreover, considering the difference between Fourier and advanced Fourier bases, 
magnitude spectra obtained by AdvFT is consistent with that by FT, as presented in Fig. \ref{fig:BFE_AMP_ByBaseFT_Compare} in Appendix 3. 

Although there is also an effect of multiple-scattering effect\cite{rehr2000modern}, 
the uncertainty in the proximity structure after 4 Å due to the effect of the multiple-scattering effect is not included in Fourier, Windowed Fourier and Advanced Fourier bases. On the other hand, the contribution to the measured data by the single-scattering effect exists [c.f. Fig. 19 \cite{rehr2000modern}]. Then, considering the limitations of our model which cannot take into account multiple-scattering effects, we shaded the area after $4$~\AA~to reduce attention to the distant coordination structure as shown in the Figs. 8, 9, 10 and 11. 

Then, by comparing magnitude spectra based on CVE optimization\cite{Akai2018}, 
we discuss the effects of prior knowledge involved in the Bayesian inference. 
Figs. \ref{fig:BFEandCVE1SE_Est_mode}(a) and (d) show 
the regression results and the obtained magnitude spectra,  
where $\lambda_{\mathrm{CVEmin}}=3.2\times10^{-5}$ and $\sigma_{\mathrm{DW}}=0.103$~\AA~are selected by CVE minimization, 
indicated with an asterisk in Fig. \ref{fig:HyperParamEst-DWsigmavslambda-CVEmap}(c). 
Although the estimated regression results are consistent with EXAFS signals from low $k$ to high $k$, as shown in Fig. \ref{fig:BFEandCVE1SE_Est_mode}(a), the optimized sparse regularization parameter $\lambda_{\mathrm{CVEmin}}=3.2\times10^{-5}$ is less than $\lambda_{\mathrm{BFEmin}}=2.3\times 10^{-2}$, 
and the number of basis functions selected at $\lambda_{\rm CVEmin}$ is $315$, which is not sparse, as shown in Fig. \ref{fig:BFEandCVE1SE_Est_mode}(d). 

To obtain a simpler (more regularized) model and avoid overfitting, the previous study \cite{Akai2018} optimized the regularization parameter $\lambda$ using the one-standard-error (1SE) rule \cite{Murphy2012}. The 1SE rule chooses the largest $\lambda$ among those whose error is within one standard error of the minimum CVE, 
and 
$\lambda_{\mathrm{CVEmin}}=2.1\times10^{-3}$ and $\sigma_{\mathrm{DW}}=0.103$~\AA~are selected, as denoted by the triangle point in Fig. \ref{fig:HyperParamEst-DWsigmavslambda-CVEmap}(c). 
Then, although CVE=$0.341$ increases by $6\%$ more than the minimum CVE, they obtained a sparse solution, with $113$ basis functions, as shown in Figs. \ref{fig:BFEandCVE1SE_Est_mode}(b) and (e). However, the 1SE rule is a heuristic method, and there is uncertainty in the regression results and selected basis functions depending on the measured data and random partitioning for the CVE. 

Let us discuss the regression results based on BFE. 
Although the estimated regression result exhibits some minor deviations for EXAFS signals at high $k<17$~\AA$^{-1}$~as shown in Fig. \ref{fig:BFEandCVE1SE_Est_mode}(c), 
magnitude spectra obtained by BFE minimization is sparse, and the number of basis functions is $82$, which is much fewer than that obtained by CVE minimization, as mentioned above. 
Although the CVE equivalently evaluates the prediction error over the whole $k$ region, 
our Bayesian approach minimizes the weighted residual sum of squares, as shown in Eq. (\ref{eq:likelihood}). 
This approach allows us to handle the $k^3$-weighted EXAFS signal appropriately while taking into account the $k^3$-weighted noise and disregarding the EXAFS signal at $k$ higher than 17.5~${\rm \AA}^{-1}$. 
Thus, although the regression model obtained by CVE minimization needs too many basis functions to match an EXAFS signals at high $k$, 
the regression model obtained by BFE minimization 
reduces the number of needed basis functions to only $82$. 
Thus, the obtained sparse magnitude spectra allows us to extract near structures more clearly. 
Moreover, 
although the magnitude spectra obtained by CVE minimization drastically increases as $R$ increases due to the effects of prior knowledge of the coordination number 
, the magnitude spectra obtained by BFE minimization is approximately proportional to $R_j^2$, as shown in Figs. \ref{fig:BFEandCVE1SE_Est_mode}(d) and (f). 

Let us consider why the Bayesian free energy (BFE) has more advantages for basis selection than the cross-validation error (CVE). As denoted by theoretical studies and practical usages \cite{anderson2004model,vrieze2012model}, the BFE selects the true model if the true model is among the candidate models considered. On the other hand, the CVE is not consistent under these circumstances \cite{vrieze2012model} and tends to select an unnecessarily large model, namely, much more bases \cite{Igarashi2018,shao1993linear}. In EXASF analysis, we use a single-scattering approximation in the theoretical background, and the appropriate model for EXAFS signals is among the candidate models considered. Thus, the magnitude spectra 
obtained by BFE minimization is a sparse and appropriate model in the EXAFS analysis for Cu and could be advantageous for other transition metals with different crystal structures. 

\subsection{Reliability of magnitude spectra based on Bayesian inference}
Our Bayesian approach enables quantitative evaluation of the reliability of the magnitude spectra.  
Although previous studies have estimated the reliability using simulation data or EXAFS models with added noise \cite{Morrison1982,Incoccia1984,Vaarkamp1998}, 
our Bayesian approach can simultaneously estimate the noise intensity included in experimental data 
and the reliability of the magnitude spectra 
under the assumption of proper and moderate prior knowledge \cite{Krappe2000,bishop2006pattern}. 
Fig. \ref{fig:VarEstConfidence}(a) shows the estimated standard deviation of the magnitude spectra, 
which is closely correlated to the coordination number, and we find that the standard deviation markedly increases 
in the long distance region ($R>6$~\AA). 
To evaluate the signal-to-noise ratio, Fig. \ref{fig:VarEstConfidence}(b) shows the magnitude spectra, 
where the error bars are the standard deviation. 
While the median ratio of the standard deviation 
to the magnitude in the intermediate range ($3<R<6$~\AA) is $37.8\%$, 
the median ratio for near structures 
($1.9 < R < 2.5$~\AA) is $18.9\%$. 
This near range corresponds to the main peak in the magnitude of the FT, 
which is due to nearest Cu-Cu single scattering. 
These results show that EXAFS can extract near structures ($R<3$~\AA) with high accuracy and structures in the intermediate range with reasonable accuracy, which is in good agreement with previous simulation studies \cite{Dalba2008,Vaarkamp1998,Morrison1982}. 
In the long distance range ($R>6$~\AA), although magnitude spectra is significant, its error increases, as shown in Fig. \ref{fig:VarEstConfidence}(b). 
Since our model is derived from a simplified model based on the singled-scattering approximation, multiple-scattering effects could be important above about $3$~\AA~from the view of EXAFS analysis by Bayesian inference. From a statistical perspective, Bayesian inference makes clear the range in application of single-scattering approximation. 

\subsection{Results of EXAFS analysis for Fe}
 \label{subsec:ResultsFe}
In this study, Figs. 8 and 9 show that Bayesian inference and AdvFT basis enables us to extract the sparse and appropriate structural parameters of copper with high accuracy in the near structures ($R<3$~\AA). This sparse solution clarify a detailed structure in the near structures, which is one of the result of our method from the physical point of view.

\begin{figure*}
  \centering
  \includegraphics[width=6.5in]{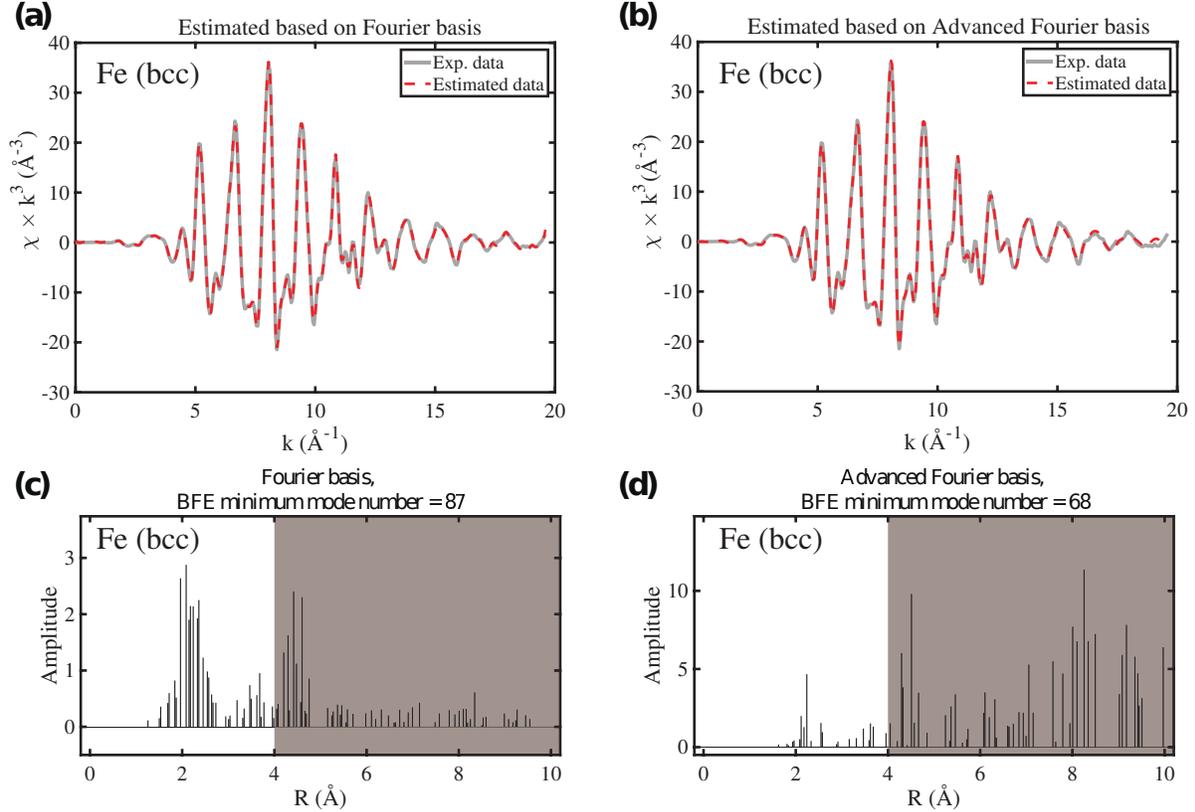}
  \caption{
Results of regression and magnitude spectrum for Fe (bcc) to compare the results with FT and AdvFT bases. (a) (b) Regression results obtained by FT and AdvFT, respectively. (c) (d) Magnitude spectra obtained by FT and AdvFT, respectively, where the $L1$-regularization parameter $\lambda$ is optimized by BFE minimization. 
    }
 \label{fig:FE_FT_AdvFT_Compare}
\end{figure*}

To clarify the important advantage over the conventional method using Fourier Basis, 
we show another result of EXAFS analysis for Fe, having different crystal structures as shown in Fig. \ref{fig:FE_FT_AdvFT_Compare}. Experimental Fe K-edge EXAFS for temperature $300$K have the body centered cubic (bcc) structures and magnitude spectra between $2$ and $4$ ~\AA~is split in two peaks \cite{rickerby1982lattice}, which is useful for testing the effectiveness of extracting the information on the near local structure \cite{timoshenko2018neural}. 

Figure \ref{fig:FE_FT_AdvFT_Compare} shows results of regression and magnitude spectrum for Fe (bcc) to compare the results by FT and by AdvFT bases. In the results, the regularization parameter and other physical parameters in AdvFT basis are optimized by Bayesian inference. 
Although the magnitude spectra obtained by FT cannot reconstruct the two peak structure from EXAFS signals, the one by AdvFT capture the structure as shown in Figs.\ref{fig:FE_FT_AdvFT_Compare} (c) and (d). 
This results are consistent with that of RMC and Neural networks 
\cite{timoshenko2018neural}. Note that our results are obtained without the knowledge of the density and chemical composition of the system and a specialized train for constructing Neural networks for each case. 
Thus, Bayesian inference and AdvFT basis efficiently extracts the sparse and appropriate structural parameters with high accuracy in the near structures. 

\section{Conclusion}
We extended Bayesian inference for EXAFS to select an appropriate basis, magnitude spectra and physical parameters from measured data alone using prior physical knowledge that can be generally applied to materials. 
To demonstrate the validity of our method, the well-known EXAFS spectrum of copper is used for EXAFS data analysis. Based on the framework of Bayesian inference and using the reliability in the coordination number, 
AdvFT is selected as an appropriate basis for the regression of EXAFS signals in a quantitative way, 
and the estimation of the Debye-Waller factor can be robustly realized by using AdvFT. 

Bayesian inference based on minimal restrictions 
allows us to not only eliminate some unphysical results but also select an appropriate basis and extract magnitude spectra and physical parameters from EXAFS signals alone, which leads to the general usage of Bayesian inference for EXAFS data analysis. 
Moreover, since the Bayesian inference framework provides flexibility in the choice of prior knowledge, combining prior information of well-known materials such as crystals obtained by FEFF signals, 
the Bayesian approach can be extended to more complex objects. 

As denoted by EXAFS signals of real data \cite{rehr2000modern,jonane2018advanced,timoshenko2020linking}, 
a complete analysis of the EXAFS signal needs to incorporate multiple scattering processes, differences in the Debye-Waller factor for the respective coordinating atoms, and the wavenumber dependence of the phase shift and mean free path. To overcome these limitations, we can extend our framework by using more appropriate bases, which include the effects calculated by FEFF \cite{Ravel2005,Newville2014} or GNAXS\cite{di2009gnxas}, and selecting the optimal bases for EXAFS signals based on Bayesian inference. 
We thus consider the extension of the model selection based on Bayesian inference to be highly flexible and applicable to basis functions beyond the single scattering approximation of EXAFS processes. Further investigation is needed to investigate the applicable scope of the Bayesian framework for EXAFS signals. 

\section*{Funding}
This work was supported by KAKENHI Grants-in-Aid for Scientific
Research on Innovative Areas (JP25120009) from the Japan Society for the
Promotion of Science (JSPS), and CREST (JPMJCR1761,JPMJCR1861) and PRESTO
(JPMJPR17N2) from the Japan Science and Technology Agency (JST). The
EXAFS measurements were conducted at the BL11 beamline in the SAGA-LS
under Proposal No. 160304I.

\bibliographystyle{aipauth4-1}
\bibliography{bibl}

\begin{thebibliography}{49}%
\makeatletter
\providecommand \@ifxundefined [1]{%
 \@ifx{#1\undefined}
}%
\providecommand \@ifnum [1]{%
 \ifnum #1\expandafter \@firstoftwo
 \else \expandafter \@secondoftwo
 \fi
}%
\providecommand \@ifx [1]{%
 \ifx #1\expandafter \@firstoftwo
 \else \expandafter \@secondoftwo
 \fi
}%
\providecommand \natexlab [1]{#1}%
\providecommand \enquote  [1]{``#1''}%
\providecommand \bibnamefont  [1]{#1}%
\providecommand \bibfnamefont [1]{#1}%
\providecommand \citenamefont [1]{#1}%
\providecommand \href@noop [0]{\@secondoftwo}%
\providecommand \href [0]{\begingroup \@sanitize@url \@href}%
\providecommand \@href[1]{\@@startlink{#1}\@@href}%
\providecommand \@@href[1]{\endgroup#1\@@endlink}%
\providecommand \@sanitize@url [0]{\catcode `\\12\catcode `\$12\catcode
  `\&12\catcode `\#12\catcode `\^12\catcode `\_12\catcode `\%12\relax}%
\providecommand \@@startlink[1]{}%
\providecommand \@@endlink[0]{}%
\providecommand \url  [0]{\begingroup\@sanitize@url \@url }%
\providecommand \@url [1]{\endgroup\@href {#1}{\urlprefix }}%
\providecommand \urlprefix  [0]{URL }%
\providecommand \Eprint [0]{\href }%
\providecommand \doibase [0]{http://dx.doi.org/}%
\providecommand \selectlanguage [0]{\@gobble}%
\providecommand \bibinfo  [0]{\@secondoftwo}%
\providecommand \bibfield  [0]{\@secondoftwo}%
\providecommand \translation [1]{[#1]}%
\providecommand \BibitemOpen [0]{}%
\providecommand \bibitemStop [0]{}%
\providecommand \bibitemNoStop [0]{.\EOS\space}%
\providecommand \EOS [0]{\spacefactor3000\relax}%
\providecommand \BibitemShut  [1]{\csname bibitem#1\endcsname}%
\let\auto@bib@innerbib\@empty
\bibitem [{\citenamefont {Akai}\ \emph {et~al.}(2018)\citenamefont {Akai},
  \citenamefont {Iwamitsu}, \citenamefont {Igarashi}, \citenamefont {Okada},
  \citenamefont {Setoyama}, \citenamefont {Okajima},\ and\ \citenamefont
  {Hirai}}]{Akai2018}%
  \BibitemOpen
  \bibfield  {author} {\bibinfo {author} {\bibnamefont {Akai}, \bibfnamefont
  {I.}}, \bibinfo {author} {\bibnamefont {Iwamitsu}, \bibfnamefont {K.}},
  \bibinfo {author} {\bibnamefont {Igarashi}, \bibfnamefont {Y.}}, \bibinfo
  {author} {\bibnamefont {Okada}, \bibfnamefont {M.}}, \bibinfo {author}
  {\bibnamefont {Setoyama}, \bibfnamefont {H.}}, \bibinfo {author}
  {\bibnamefont {Okajima}, \bibfnamefont {T.}}, \ and\ \bibinfo {author}
  {\bibnamefont {Hirai}, \bibfnamefont {Y.}},\ }\href@noop {} {\bibfield
  {journal} {\bibinfo  {journal} {Journal of the Physical Society of Japan}\
  }\textbf {\bibinfo {volume} {87}},\ \bibinfo {pages} {074003} (\bibinfo
  {year} {2018})}\BibitemShut {NoStop}%
\bibitem [{\citenamefont {Anderson}\ and\ \citenamefont
  {Burnham}(2004)}]{anderson2004model}%
  \BibitemOpen
  \bibfield  {author} {\bibinfo {author} {\bibnamefont {Anderson},
  \bibfnamefont {D.}}\ and\ \bibinfo {author} {\bibnamefont {Burnham},
  \bibfnamefont {K.}},\ }\href@noop {} {\bibfield  {journal} {\bibinfo
  {journal} {Second. NY: Springer-Verlag}\ }\textbf {\bibinfo {volume} {63}}
  (\bibinfo {year} {2004})}\BibitemShut {NoStop}%
\bibitem [{\citenamefont {Bishop}(2006)}]{bishop2006pattern}%
  \BibitemOpen
  \bibfield  {author} {\bibinfo {author} {\bibnamefont {Bishop}, \bibfnamefont
  {C.~M.}},\ }\href@noop {} {\emph {\bibinfo {title} {Pattern recognition and
  machine learning}}}\ (\bibinfo  {publisher} {springer},\ \bibinfo {year}
  {2006})\BibitemShut {NoStop}%
\bibitem [{\citenamefont {Bunker}(2010)}]{Bunker2010}%
  \BibitemOpen
  \bibfield  {author} {\bibinfo {author} {\bibnamefont {Bunker}, \bibfnamefont
  {G.}},\ }\href@noop {} {\emph {\bibinfo {title} {Introduction to XAFS: a
  practical guide to X-ray absorption fine structure spectroscopy}}}\ (\bibinfo
   {publisher} {Cambridge University Press},\ \bibinfo {year}
  {2010})\BibitemShut {NoStop}%
\bibitem [{\citenamefont {Calvin}(2013)}]{Calvin2013}%
  \BibitemOpen
  \bibfield  {author} {\bibinfo {author} {\bibnamefont {Calvin}, \bibfnamefont
  {S.}},\ }\href@noop {} {\emph {\bibinfo {title} {XAFS for Everyone}}}\
  (\bibinfo  {publisher} {CRC press},\ \bibinfo {year} {2013})\BibitemShut
  {NoStop}%
\bibitem [{\citenamefont {Cover}\ and\ \citenamefont
  {Van~Campenhout}(1977)}]{Cover1977}%
  \BibitemOpen
  \bibfield  {author} {\bibinfo {author} {\bibnamefont {Cover}, \bibfnamefont
  {T.~M.}}\ and\ \bibinfo {author} {\bibnamefont {Van~Campenhout},
  \bibfnamefont {J.~M.}},\ }\href@noop {} {\bibfield  {journal} {\bibinfo
  {journal} {IEEE transactions on systems, man, and cybernetics}\ }\textbf
  {\bibinfo {volume} {7}},\ \bibinfo {pages} {657} (\bibinfo {year}
  {1977})}\BibitemShut {NoStop}%
\bibitem [{\citenamefont {Dalba}, \citenamefont {Afify},\ and\ \citenamefont
  {Rocca}(2008)}]{Dalba2008}%
  \BibitemOpen
  \bibfield  {author} {\bibinfo {author} {\bibnamefont {Dalba}, \bibfnamefont
  {G.}}, \bibinfo {author} {\bibnamefont {Afify}, \bibfnamefont {N.~D.}}, \
  and\ \bibinfo {author} {\bibnamefont {Rocca}, \bibfnamefont {F.}},\
  }\href@noop {} {\bibfield  {journal} {\bibinfo  {journal} {Physics and
  Chemistry of Glasses-European Journal of Glass Science and Technology Part
  B}\ }\textbf {\bibinfo {volume} {49}},\ \bibinfo {pages} {149} (\bibinfo
  {year} {2008})}\BibitemShut {NoStop}%
\bibitem [{\citenamefont {Di~Cicco}(2009)}]{di2009gnxas}%
  \BibitemOpen
  \bibfield  {author} {\bibinfo {author} {\bibnamefont {Di~Cicco},
  \bibfnamefont {A.}},\ }\href@noop {} {\bibfield  {journal} {\bibinfo
  {journal} {Extended suite of programs for advanced X-ray absorption
  data-analysis: methodology and practice. Task, Gdansk}\ } (\bibinfo {year}
  {2009})}\BibitemShut {NoStop}%
\bibitem [{\citenamefont {Efron}\ \emph {et~al.}(2004)\citenamefont {Efron},
  \citenamefont {Hastie}, \citenamefont {Johnstone}, \citenamefont {Tibshirani}
  \emph {et~al.}}]{Efron2004}%
  \BibitemOpen
  \bibfield  {author} {\bibinfo {author} {\bibnamefont {Efron}, \bibfnamefont
  {B.}}, \bibinfo {author} {\bibnamefont {Hastie}, \bibfnamefont {T.}},
  \bibinfo {author} {\bibnamefont {Johnstone}, \bibfnamefont {I.}}, \bibinfo
  {author} {\bibnamefont {Tibshirani}, \bibfnamefont {R.}},  \emph {et~al.},\
  }\href@noop {} {\bibfield  {journal} {\bibinfo  {journal} {The Annals of
  statistics}\ }\textbf {\bibinfo {volume} {32}},\ \bibinfo {pages} {407}
  (\bibinfo {year} {2004})}\BibitemShut {NoStop}%
\bibitem [{\citenamefont {Ershov}\ \emph {et~al.}(1981)\citenamefont {Ershov},
  \citenamefont {Ageev}, \citenamefont {Vasin},\ and\ \citenamefont
  {Babanov}}]{ershov1981new}%
  \BibitemOpen
  \bibfield  {author} {\bibinfo {author} {\bibnamefont {Ershov}, \bibfnamefont
  {N.}}, \bibinfo {author} {\bibnamefont {Ageev}, \bibfnamefont {A.}}, \bibinfo
  {author} {\bibnamefont {Vasin}, \bibfnamefont {V.}}, \ and\ \bibinfo {author}
  {\bibnamefont {Babanov}, \bibfnamefont {Y.~A.}},\ }\href@noop {} {\bibfield
  {journal} {\bibinfo  {journal} {physica status solidi (b)}\ }\textbf
  {\bibinfo {volume} {108}},\ \bibinfo {pages} {103} (\bibinfo {year}
  {1981})}\BibitemShut {NoStop}%
\bibitem [{\citenamefont {Fornasini}\ \emph {et~al.}(2004)\citenamefont
  {Fornasini}, \citenamefont {a~Beccara}, \citenamefont {Dalba}, \citenamefont
  {Grisenti}, \citenamefont {Sanson}, \citenamefont {Vaccari},\ and\
  \citenamefont {Rocca}}]{FornasiniPhysRevB2004}%
  \BibitemOpen
  \bibfield  {author} {\bibinfo {author} {\bibnamefont {Fornasini},
  \bibfnamefont {P.}}, \bibinfo {author} {\bibnamefont {a~Beccara},
  \bibfnamefont {S.}}, \bibinfo {author} {\bibnamefont {Dalba}, \bibfnamefont
  {G.}}, \bibinfo {author} {\bibnamefont {Grisenti}, \bibfnamefont {R.}},
  \bibinfo {author} {\bibnamefont {Sanson}, \bibfnamefont {A.}}, \bibinfo
  {author} {\bibnamefont {Vaccari}, \bibfnamefont {M.}}, \ and\ \bibinfo
  {author} {\bibnamefont {Rocca}, \bibfnamefont {F.}},\ }\href {\doibase
  10.1103/PhysRevB.70.174301} {\bibfield  {journal} {\bibinfo  {journal} {Phys.
  Rev. B}\ }\textbf {\bibinfo {volume} {70}},\ \bibinfo {pages} {174301}
  (\bibinfo {year} {2004})}\BibitemShut {NoStop}%
\bibitem [{\citenamefont {Funke}, \citenamefont {Chukalina},\ and\
  \citenamefont {Scheinost}(2007)}]{funke2007new}%
  \BibitemOpen
  \bibfield  {author} {\bibinfo {author} {\bibnamefont {Funke}, \bibfnamefont
  {H.}}, \bibinfo {author} {\bibnamefont {Chukalina}, \bibfnamefont {M.}}, \
  and\ \bibinfo {author} {\bibnamefont {Scheinost}, \bibfnamefont {A.~C.}},\
  }\href@noop {} {\bibfield  {journal} {\bibinfo  {journal} {Journal of
  synchrotron radiation}\ }\textbf {\bibinfo {volume} {14}},\ \bibinfo {pages}
  {426} (\bibinfo {year} {2007})}\BibitemShut {NoStop}%
\bibitem [{\citenamefont {Funke}, \citenamefont {Scheinost},\ and\
  \citenamefont {Chukalina}(2005)}]{funke2005wavelet}%
  \BibitemOpen
  \bibfield  {author} {\bibinfo {author} {\bibnamefont {Funke}, \bibfnamefont
  {H.}}, \bibinfo {author} {\bibnamefont {Scheinost}, \bibfnamefont {A.}}, \
  and\ \bibinfo {author} {\bibnamefont {Chukalina}, \bibfnamefont {M.}},\
  }\href@noop {} {\bibfield  {journal} {\bibinfo  {journal} {Physical Review
  B}\ }\textbf {\bibinfo {volume} {71}},\ \bibinfo {pages} {094110} (\bibinfo
  {year} {2005})}\BibitemShut {NoStop}%
\bibitem [{\citenamefont {Grossmann}, \citenamefont {Kronland-Martinet},\ and\
  \citenamefont {Morlet}(1990)}]{grossmann1990reading}%
  \BibitemOpen
  \bibfield  {author} {\bibinfo {author} {\bibnamefont {Grossmann},
  \bibfnamefont {A.}}, \bibinfo {author} {\bibnamefont {Kronland-Martinet},
  \bibfnamefont {R.}}, \ and\ \bibinfo {author} {\bibnamefont {Morlet},
  \bibfnamefont {J.}},\ }in\ \href@noop {} {\emph {\bibinfo {booktitle}
  {Wavelets}}}\ (\bibinfo  {publisher} {Springer},\ \bibinfo {year} {1990})\
  pp.\ \bibinfo {pages} {2--20}\BibitemShut {NoStop}%
\bibitem [{\citenamefont {Hafner}\ \emph {et~al.}(2018)\citenamefont {Hafner},
  \citenamefont {Tran}, \citenamefont {Lillicrap}, \citenamefont {Irpan},\ and\
  \citenamefont {Davidson}}]{hafner2018reliable}%
  \BibitemOpen
  \bibfield  {author} {\bibinfo {author} {\bibnamefont {Hafner}, \bibfnamefont
  {D.}}, \bibinfo {author} {\bibnamefont {Tran}, \bibfnamefont {D.}}, \bibinfo
  {author} {\bibnamefont {Lillicrap}, \bibfnamefont {T.}}, \bibinfo {author}
  {\bibnamefont {Irpan}, \bibfnamefont {A.}}, \ and\ \bibinfo {author}
  {\bibnamefont {Davidson}, \bibfnamefont {J.}},\ }\href@noop {} {\  (\bibinfo
  {year} {2018})}\BibitemShut {NoStop}%
\bibitem [{\citenamefont {Hermann}(2017)}]{hermann2017crystallography}%
  \BibitemOpen
  \bibfield  {author} {\bibinfo {author} {\bibnamefont {Hermann}, \bibfnamefont
  {K.}},\ }\href@noop {} {\emph {\bibinfo {title} {Crystallography and Surface
  Structure: An Introduction for Surface Scientists and Nanoscientists}}}\
  (\bibinfo  {publisher} {John Wiley \& Sons},\ \bibinfo {year}
  {2017})\BibitemShut {NoStop}%
\bibitem [{\citenamefont {Iesari}\ \emph {et~al.}(2018)\citenamefont {Iesari},
  \citenamefont {Hatada}, \citenamefont {Trapananti}, \citenamefont
  {Minicucci},\ and\ \citenamefont {Di~Cicco}}]{iesari2018gnxas}%
  \BibitemOpen
  \bibfield  {author} {\bibinfo {author} {\bibnamefont {Iesari}, \bibfnamefont
  {F.}}, \bibinfo {author} {\bibnamefont {Hatada}, \bibfnamefont {K.}},
  \bibinfo {author} {\bibnamefont {Trapananti}, \bibfnamefont {A.}}, \bibinfo
  {author} {\bibnamefont {Minicucci}, \bibfnamefont {M.}}, \ and\ \bibinfo
  {author} {\bibnamefont {Di~Cicco}, \bibfnamefont {A.}},\ }in\ \href@noop {}
  {\emph {\bibinfo {booktitle} {Multiple Scattering Theory for
  Spectroscopies}}}\ (\bibinfo  {publisher} {Springer},\ \bibinfo {year}
  {2018})\ pp.\ \bibinfo {pages} {221--256}\BibitemShut {NoStop}%
\bibitem [{\citenamefont {Igarashi}\ \emph {et~al.}(2018)\citenamefont
  {Igarashi}, \citenamefont {Takenaka}, \citenamefont {Nakanishi-Ohno},
  \citenamefont {Uemura}, \citenamefont {Ikeda},\ and\ \citenamefont
  {Okada}}]{Igarashi2018}%
  \BibitemOpen
  \bibfield  {author} {\bibinfo {author} {\bibnamefont {Igarashi},
  \bibfnamefont {Y.}}, \bibinfo {author} {\bibnamefont {Takenaka},
  \bibfnamefont {H.}}, \bibinfo {author} {\bibnamefont {Nakanishi-Ohno},
  \bibfnamefont {Y.}}, \bibinfo {author} {\bibnamefont {Uemura}, \bibfnamefont
  {M.}}, \bibinfo {author} {\bibnamefont {Ikeda}, \bibfnamefont {S.}}, \ and\
  \bibinfo {author} {\bibnamefont {Okada}, \bibfnamefont {M.}},\ }\href@noop {}
  {\bibfield  {journal} {\bibinfo  {journal} {Journal of the Physical Society
  of Japan}\ }\textbf {\bibinfo {volume} {87}},\ \bibinfo {pages} {044802}
  (\bibinfo {year} {2018})}\BibitemShut {NoStop}%
\bibitem [{\citenamefont {Incoccia}\ and\ \citenamefont
  {Mobilio}(1984)}]{Incoccia1984}%
  \BibitemOpen
  \bibfield  {author} {\bibinfo {author} {\bibnamefont {Incoccia},
  \bibfnamefont {L.}}\ and\ \bibinfo {author} {\bibnamefont {Mobilio},
  \bibfnamefont {S.}},\ }\href@noop {} {\bibfield  {journal} {\bibinfo
  {journal} {Il Nuovo Cimento D}\ }\textbf {\bibinfo {volume} {3}},\ \bibinfo
  {pages} {867} (\bibinfo {year} {1984})}\BibitemShut {NoStop}%
\bibitem [{\citenamefont {Jonane}, \citenamefont {Anspoks},\ and\ \citenamefont
  {Kuzmin}(2018)}]{jonane2018advanced}%
  \BibitemOpen
  \bibfield  {author} {\bibinfo {author} {\bibnamefont {Jonane}, \bibfnamefont
  {I.}}, \bibinfo {author} {\bibnamefont {Anspoks}, \bibfnamefont {A.}}, \ and\
  \bibinfo {author} {\bibnamefont {Kuzmin}, \bibfnamefont {A.}},\ }\href@noop
  {} {\bibfield  {journal} {\bibinfo  {journal} {Modelling and Simulation in
  Materials Science and Engineering}\ }\textbf {\bibinfo {volume} {26}},\
  \bibinfo {pages} {025004} (\bibinfo {year} {2018})}\BibitemShut {NoStop}%
\bibitem [{\citenamefont {Krappe}\ and\ \citenamefont
  {Rossner}(2000)}]{Krappe2000}%
  \BibitemOpen
  \bibfield  {author} {\bibinfo {author} {\bibnamefont {Krappe}, \bibfnamefont
  {H.}}\ and\ \bibinfo {author} {\bibnamefont {Rossner}, \bibfnamefont {H.}},\
  }\href@noop {} {\bibfield  {journal} {\bibinfo  {journal} {Physical Review
  B}\ }\textbf {\bibinfo {volume} {61}},\ \bibinfo {pages} {6596} (\bibinfo
  {year} {2000})}\BibitemShut {NoStop}%
\bibitem [{\citenamefont {Krappe}\ and\ \citenamefont
  {Rossner}(2002)}]{Krappe2002}%
  \BibitemOpen
  \bibfield  {author} {\bibinfo {author} {\bibnamefont {Krappe}, \bibfnamefont
  {H.}}\ and\ \bibinfo {author} {\bibnamefont {Rossner}, \bibfnamefont {H.}},\
  }\href@noop {} {\bibfield  {journal} {\bibinfo  {journal} {Physical Review
  B}\ }\textbf {\bibinfo {volume} {66}},\ \bibinfo {pages} {184303} (\bibinfo
  {year} {2002})}\BibitemShut {NoStop}%
\bibitem [{\citenamefont {Kunicke}\ \emph {et~al.}(2005)\citenamefont
  {Kunicke}, \citenamefont {Kamensky}, \citenamefont {Babanov},\ and\
  \citenamefont {Funke}}]{kunicke2005efficient}%
  \BibitemOpen
  \bibfield  {author} {\bibinfo {author} {\bibnamefont {Kunicke}, \bibfnamefont
  {M.}}, \bibinfo {author} {\bibnamefont {Kamensky}, \bibfnamefont {I.~Y.}},
  \bibinfo {author} {\bibnamefont {Babanov}, \bibfnamefont {Y.~A.}}, \ and\
  \bibinfo {author} {\bibnamefont {Funke}, \bibfnamefont {H.}},\ }\href@noop {}
  {\bibfield  {journal} {\bibinfo  {journal} {Physica Scripta}\ }\textbf
  {\bibinfo {volume} {2005}},\ \bibinfo {pages} {237} (\bibinfo {year}
  {2005})}\BibitemShut {NoStop}%
\bibitem [{\citenamefont {Kuzmin}\ and\ \citenamefont
  {Purans}(2000)}]{kuzmin2000dehydration}%
  \BibitemOpen
  \bibfield  {author} {\bibinfo {author} {\bibnamefont {Kuzmin}, \bibfnamefont
  {A.}}\ and\ \bibinfo {author} {\bibnamefont {Purans}, \bibfnamefont {J.}},\
  }\href@noop {} {\bibfield  {journal} {\bibinfo  {journal} {Journal of
  Physics: Condensed Matter}\ }\textbf {\bibinfo {volume} {12}},\ \bibinfo
  {pages} {1959} (\bibinfo {year} {2000})}\BibitemShut {NoStop}%
\bibitem [{\citenamefont {Martini}\ \emph {et~al.}(2017)\citenamefont
  {Martini}, \citenamefont {Borfecchia}, \citenamefont {Lomachenko},
  \citenamefont {Pankin}, \citenamefont {Negri}, \citenamefont {Berlier},
  \citenamefont {Beato}, \citenamefont {Falsig}, \citenamefont {Bordiga},\ and\
  \citenamefont {Lamberti}}]{martini2017composition}%
  \BibitemOpen
  \bibfield  {author} {\bibinfo {author} {\bibnamefont {Martini}, \bibfnamefont
  {A.}}, \bibinfo {author} {\bibnamefont {Borfecchia}, \bibfnamefont {E.}},
  \bibinfo {author} {\bibnamefont {Lomachenko}, \bibfnamefont {K.}}, \bibinfo
  {author} {\bibnamefont {Pankin}, \bibfnamefont {I.}}, \bibinfo {author}
  {\bibnamefont {Negri}, \bibfnamefont {C.}}, \bibinfo {author} {\bibnamefont
  {Berlier}, \bibfnamefont {G.}}, \bibinfo {author} {\bibnamefont {Beato},
  \bibfnamefont {P.}}, \bibinfo {author} {\bibnamefont {Falsig}, \bibfnamefont
  {H.}}, \bibinfo {author} {\bibnamefont {Bordiga}, \bibfnamefont {S.}}, \ and\
  \bibinfo {author} {\bibnamefont {Lamberti}, \bibfnamefont {C.}},\ }\href@noop
  {} {\bibfield  {journal} {\bibinfo  {journal} {Chemical science}\ }\textbf
  {\bibinfo {volume} {8}},\ \bibinfo {pages} {6836} (\bibinfo {year}
  {2017})}\BibitemShut {NoStop}%
\bibitem [{\citenamefont {Morrison}, \citenamefont {Shenoy},\ and\
  \citenamefont {Niarchos}(1982)}]{Morrison1982}%
  \BibitemOpen
  \bibfield  {author} {\bibinfo {author} {\bibnamefont {Morrison},
  \bibfnamefont {T.~I.}}, \bibinfo {author} {\bibnamefont {Shenoy},
  \bibfnamefont {G.}}, \ and\ \bibinfo {author} {\bibnamefont {Niarchos},
  \bibfnamefont {D.}},\ }\href@noop {} {\bibfield  {journal} {\bibinfo
  {journal} {Journal of Applied Crystallography}\ }\textbf {\bibinfo {volume}
  {15}},\ \bibinfo {pages} {388} (\bibinfo {year} {1982})}\BibitemShut
  {NoStop}%
\bibitem [{\citenamefont {Mototake}\ \emph {et~al.}(2018)\citenamefont
  {Mototake}, \citenamefont {Igarashi}, \citenamefont {Takenaka}, \citenamefont
  {Nagata},\ and\ \citenamefont {Okada}}]{Mototake2018}%
  \BibitemOpen
  \bibfield  {author} {\bibinfo {author} {\bibnamefont {Mototake},
  \bibfnamefont {Y.}}, \bibinfo {author} {\bibnamefont {Igarashi},
  \bibfnamefont {Y.}}, \bibinfo {author} {\bibnamefont {Takenaka},
  \bibfnamefont {H.}}, \bibinfo {author} {\bibnamefont {Nagata}, \bibfnamefont
  {K.}}, \ and\ \bibinfo {author} {\bibnamefont {Okada}, \bibfnamefont {M.}},\
  }\href@noop {} {\bibfield  {journal} {\bibinfo  {journal} {Journal of the
  Physical Society of Japan}\ }\textbf {\bibinfo {volume} {87}},\ \bibinfo
  {pages} {114004} (\bibinfo {year} {2018})}\BibitemShut {NoStop}%
\bibitem [{\citenamefont {Murphy}(2012)}]{Murphy2012}%
  \BibitemOpen
  \bibfield  {author} {\bibinfo {author} {\bibnamefont {Murphy}, \bibfnamefont
  {K.~P.}},\ }\href@noop {} {\enquote {\bibinfo {title} {Machine learning: A
  probabilistic perspective. adaptive computation and machine learning},}\ }
  (\bibinfo {year} {2012})\BibitemShut {NoStop}%
\bibitem [{\citenamefont {Newville}(2014)}]{Newville2014}%
  \BibitemOpen
  \bibfield  {author} {\bibinfo {author} {\bibnamefont {Newville},
  \bibfnamefont {M.}},\ }\href@noop {} {\bibfield  {journal} {\bibinfo
  {journal} {Reviews in Mineralogy and Geochemistry}\ }\textbf {\bibinfo
  {volume} {78}},\ \bibinfo {pages} {33} (\bibinfo {year} {2014})}\BibitemShut
  {NoStop}%
\bibitem [{\citenamefont {Okajima}\ \emph {et~al.}(2013)\citenamefont
  {Okajima}, \citenamefont {Sumitani}, \citenamefont {Kawamoto},\ and\
  \citenamefont {Kobayashi}}]{Okajima2013}%
  \BibitemOpen
  \bibfield  {author} {\bibinfo {author} {\bibnamefont {Okajima}, \bibfnamefont
  {T.}}, \bibinfo {author} {\bibnamefont {Sumitani}, \bibfnamefont {K.}},
  \bibinfo {author} {\bibnamefont {Kawamoto}, \bibfnamefont {M.}}, \ and\
  \bibinfo {author} {\bibnamefont {Kobayashi}, \bibfnamefont {E.}},\ }in\
  \href@noop {} {\emph {\bibinfo {booktitle} {Journal of Physics: Conference
  Series}}},\ Vol.\ \bibinfo {volume} {430}\ (\bibinfo {organization} {IOP
  Publishing},\ \bibinfo {year} {2013})\ p.\ \bibinfo {pages}
  {012088}\BibitemShut {NoStop}%
\bibitem [{\citenamefont {Palmer}, \citenamefont {Pfund},\ and\ \citenamefont
  {Fulton}(1996)}]{palmer1996direct}%
  \BibitemOpen
  \bibfield  {author} {\bibinfo {author} {\bibnamefont {Palmer}, \bibfnamefont
  {B.~J.}}, \bibinfo {author} {\bibnamefont {Pfund}, \bibfnamefont {D.~M.}}, \
  and\ \bibinfo {author} {\bibnamefont {Fulton}, \bibfnamefont {J.~L.}},\
  }\href@noop {} {\bibfield  {journal} {\bibinfo  {journal} {The Journal of
  Physical Chemistry}\ }\textbf {\bibinfo {volume} {100}},\ \bibinfo {pages}
  {13393} (\bibinfo {year} {1996})}\BibitemShut {NoStop}%
\bibitem [{\citenamefont {Ravel}\ and\ \citenamefont
  {Newville}(2005)}]{Ravel2005}%
  \BibitemOpen
  \bibfield  {author} {\bibinfo {author} {\bibnamefont {Ravel}, \bibfnamefont
  {B.}}\ and\ \bibinfo {author} {\bibnamefont {Newville}, \bibfnamefont {M.}},\
  }\href@noop {} {\bibfield  {journal} {\bibinfo  {journal} {Journal of
  synchrotron radiation}\ }\textbf {\bibinfo {volume} {12}},\ \bibinfo {pages}
  {537} (\bibinfo {year} {2005})}\BibitemShut {NoStop}%
\bibitem [{\citenamefont {Rehr}\ and\ \citenamefont
  {Albers}(2000)}]{rehr2000modern}%
  \BibitemOpen
  \bibfield  {author} {\bibinfo {author} {\bibnamefont {Rehr}, \bibfnamefont
  {J.}}\ and\ \bibinfo {author} {\bibnamefont {Albers}, \bibfnamefont {R.}},\
  }\href@noop {} {\bibfield  {journal} {\bibinfo  {journal} {Rev. Mod. Phys}\
  }\textbf {\bibinfo {volume} {72}},\ \bibinfo {pages} {621} (\bibinfo {year}
  {2000})}\BibitemShut {NoStop}%
\bibitem [{\citenamefont {Rickerby}(1982)}]{rickerby1982lattice}%
  \BibitemOpen
  \bibfield  {author} {\bibinfo {author} {\bibnamefont {Rickerby},
  \bibfnamefont {D.}},\ }\href@noop {} {\bibfield  {journal} {\bibinfo
  {journal} {Metal Science}\ }\textbf {\bibinfo {volume} {16}},\ \bibinfo
  {pages} {495} (\bibinfo {year} {1982})}\BibitemShut {NoStop}%
\bibitem [{\citenamefont {Sayers}, \citenamefont {Stern},\ and\ \citenamefont
  {Lytle}(1971)}]{sayers1971new}%
  \BibitemOpen
  \bibfield  {author} {\bibinfo {author} {\bibnamefont {Sayers}, \bibfnamefont
  {D.~E.}}, \bibinfo {author} {\bibnamefont {Stern}, \bibfnamefont {E.~A.}}, \
  and\ \bibinfo {author} {\bibnamefont {Lytle}, \bibfnamefont {F.~W.}},\
  }\href@noop {} {\bibfield  {journal} {\bibinfo  {journal} {Physical review
  letters}\ }\textbf {\bibinfo {volume} {27}},\ \bibinfo {pages} {1204}
  (\bibinfo {year} {1971})}\BibitemShut {NoStop}%
\bibitem [{\citenamefont {Shao}(1993)}]{shao1993linear}%
  \BibitemOpen
  \bibfield  {author} {\bibinfo {author} {\bibnamefont {Shao}, \bibfnamefont
  {J.}},\ }\href@noop {} {\bibfield  {journal} {\bibinfo  {journal} {Journal of
  the American statistical Association}\ }\textbf {\bibinfo {volume} {88}},\
  \bibinfo {pages} {486} (\bibinfo {year} {1993})}\BibitemShut {NoStop}%
\bibitem [{\citenamefont {Snoek}, \citenamefont {Larochelle},\ and\
  \citenamefont {Adams}(2012)}]{snoek2012practical}%
  \BibitemOpen
  \bibfield  {author} {\bibinfo {author} {\bibnamefont {Snoek}, \bibfnamefont
  {J.}}, \bibinfo {author} {\bibnamefont {Larochelle}, \bibfnamefont {H.}}, \
  and\ \bibinfo {author} {\bibnamefont {Adams}, \bibfnamefont {R.~P.}},\ }in\
  \href@noop {} {\emph {\bibinfo {booktitle} {Advances in neural information
  processing systems}}}\ (\bibinfo {year} {2012})\ pp.\ \bibinfo {pages}
  {2951--2959}\BibitemShut {NoStop}%
\bibitem [{\citenamefont {Stern}\ \emph {et~al.}(1995)\citenamefont {Stern},
  \citenamefont {Newville}, \citenamefont {Ravel}, \citenamefont {Yacoby},\
  and\ \citenamefont {Haskel}}]{Stern1995}%
  \BibitemOpen
  \bibfield  {author} {\bibinfo {author} {\bibnamefont {Stern}, \bibfnamefont
  {E.}}, \bibinfo {author} {\bibnamefont {Newville}, \bibfnamefont {M.}},
  \bibinfo {author} {\bibnamefont {Ravel}, \bibfnamefont {B.}}, \bibinfo
  {author} {\bibnamefont {Yacoby}, \bibfnamefont {Y.}}, \ and\ \bibinfo
  {author} {\bibnamefont {Haskel}, \bibfnamefont {D.}},\ }\href@noop {}
  {\bibfield  {journal} {\bibinfo  {journal} {Physica B: Condensed Matter}\
  }\textbf {\bibinfo {volume} {208}},\ \bibinfo {pages} {117} (\bibinfo {year}
  {1995})}\BibitemShut {NoStop}%
\bibitem [{\citenamefont {Stern}, \citenamefont {Sayers},\ and\ \citenamefont
  {Lytle}(1975)}]{SternPhysRevB1975}%
  \BibitemOpen
  \bibfield  {author} {\bibinfo {author} {\bibnamefont {Stern}, \bibfnamefont
  {E.~A.}}, \bibinfo {author} {\bibnamefont {Sayers}, \bibfnamefont {D.~E.}}, \
  and\ \bibinfo {author} {\bibnamefont {Lytle}, \bibfnamefont {F.~W.}},\ }\href
  {\doibase 10.1103/PhysRevB.11.4836} {\bibfield  {journal} {\bibinfo
  {journal} {Phys. Rev. B}\ }\textbf {\bibinfo {volume} {11}},\ \bibinfo
  {pages} {4836} (\bibinfo {year} {1975})}\BibitemShut {NoStop}%
\bibitem [{\citenamefont {Teo}(2012)}]{Teo2012}%
  \BibitemOpen
  \bibfield  {author} {\bibinfo {author} {\bibnamefont {Teo}, \bibfnamefont
  {B.~K.}},\ }\href@noop {} {\emph {\bibinfo {title} {EXAFS: basic principles
  and data analysis}}},\ Vol.~\bibinfo {volume} {9}\ (\bibinfo  {publisher}
  {Springer Science \& Business Media},\ \bibinfo {year} {2012})\BibitemShut
  {NoStop}%
\bibitem [{\citenamefont {Tibshirani}(1996)}]{Tibshirani1996}%
  \BibitemOpen
  \bibfield  {author} {\bibinfo {author} {\bibnamefont {Tibshirani},
  \bibfnamefont {R.}},\ }\href@noop {} {\bibfield  {journal} {\bibinfo
  {journal} {Journal of the Royal Statistical Society. Series B
  (Methodological)}\ ,\ \bibinfo {pages} {267}} (\bibinfo {year}
  {1996})}\BibitemShut {NoStop}%
\bibitem [{\citenamefont {Timoshenko}\ \emph
  {et~al.}(2018{\natexlab{a}})\citenamefont {Timoshenko}, \citenamefont
  {Anspoks}, \citenamefont {Cintins}, \citenamefont {Kuzmin}, \citenamefont
  {Purans},\ and\ \citenamefont {Frenkel}}]{timoshenko2018neural}%
  \BibitemOpen
  \bibfield  {author} {\bibinfo {author} {\bibnamefont {Timoshenko},
  \bibfnamefont {J.}}, \bibinfo {author} {\bibnamefont {Anspoks}, \bibfnamefont
  {A.}}, \bibinfo {author} {\bibnamefont {Cintins}, \bibfnamefont {A.}},
  \bibinfo {author} {\bibnamefont {Kuzmin}, \bibfnamefont {A.}}, \bibinfo
  {author} {\bibnamefont {Purans}, \bibfnamefont {J.}}, \ and\ \bibinfo
  {author} {\bibnamefont {Frenkel}, \bibfnamefont {A.~I.}},\ }\href@noop {}
  {\bibfield  {journal} {\bibinfo  {journal} {Physical review letters}\
  }\textbf {\bibinfo {volume} {120}},\ \bibinfo {pages} {225502} (\bibinfo
  {year} {2018}{\natexlab{a}})}\BibitemShut {NoStop}%
\bibitem [{\citenamefont {Timoshenko}\ \emph {et~al.}(2020)\citenamefont
  {Timoshenko}, \citenamefont {Jeon}, \citenamefont {Sinev}, \citenamefont
  {Haase}, \citenamefont {Herzog},\ and\ \citenamefont
  {Cuenya}}]{timoshenko2020linking}%
  \BibitemOpen
  \bibfield  {author} {\bibinfo {author} {\bibnamefont {Timoshenko},
  \bibfnamefont {J.}}, \bibinfo {author} {\bibnamefont {Jeon}, \bibfnamefont
  {H.~S.}}, \bibinfo {author} {\bibnamefont {Sinev}, \bibfnamefont {I.}},
  \bibinfo {author} {\bibnamefont {Haase}, \bibfnamefont {F.~T.}}, \bibinfo
  {author} {\bibnamefont {Herzog}, \bibfnamefont {A.}}, \ and\ \bibinfo
  {author} {\bibnamefont {Cuenya}, \bibfnamefont {B.~R.}},\ }\href@noop {}
  {\bibfield  {journal} {\bibinfo  {journal} {Chemical Science}\ }\textbf
  {\bibinfo {volume} {11}},\ \bibinfo {pages} {3727} (\bibinfo {year}
  {2020})}\BibitemShut {NoStop}%
\bibitem [{\citenamefont {Timoshenko}\ and\ \citenamefont
  {Kuzmin}(2009)}]{timoshenko2009wavelet}%
  \BibitemOpen
  \bibfield  {author} {\bibinfo {author} {\bibnamefont {Timoshenko},
  \bibfnamefont {J.}}\ and\ \bibinfo {author} {\bibnamefont {Kuzmin},
  \bibfnamefont {A.}},\ }\href@noop {} {\bibfield  {journal} {\bibinfo
  {journal} {Computer Physics Communications}\ }\textbf {\bibinfo {volume}
  {180}},\ \bibinfo {pages} {920} (\bibinfo {year} {2009})}\BibitemShut
  {NoStop}%
\bibitem [{\citenamefont {Timoshenko}, \citenamefont {Kuzmin},\ and\
  \citenamefont {Purans}(2012)}]{timoshenko2012reverse}%
  \BibitemOpen
  \bibfield  {author} {\bibinfo {author} {\bibnamefont {Timoshenko},
  \bibfnamefont {J.}}, \bibinfo {author} {\bibnamefont {Kuzmin}, \bibfnamefont
  {A.}}, \ and\ \bibinfo {author} {\bibnamefont {Purans}, \bibfnamefont {J.}},\
  }\href@noop {} {\bibfield  {journal} {\bibinfo  {journal} {Computer Physics
  Communications}\ }\textbf {\bibinfo {volume} {183}},\ \bibinfo {pages} {1237}
  (\bibinfo {year} {2012})}\BibitemShut {NoStop}%
\bibitem [{\citenamefont {Timoshenko}, \citenamefont {Kuzmin},\ and\
  \citenamefont {Purans}(2014)}]{timoshenko2014exafs}%
  \BibitemOpen
  \bibfield  {author} {\bibinfo {author} {\bibnamefont {Timoshenko},
  \bibfnamefont {J.}}, \bibinfo {author} {\bibnamefont {Kuzmin}, \bibfnamefont
  {A.}}, \ and\ \bibinfo {author} {\bibnamefont {Purans}, \bibfnamefont {J.}},\
  }\href@noop {} {\bibfield  {journal} {\bibinfo  {journal} {Journal of
  Physics: Condensed Matter}\ }\textbf {\bibinfo {volume} {26}},\ \bibinfo
  {pages} {055401} (\bibinfo {year} {2014})}\BibitemShut {NoStop}%
\bibitem [{\citenamefont {Timoshenko}\ \emph
  {et~al.}(2018{\natexlab{b}})\citenamefont {Timoshenko}, \citenamefont
  {Wrasman}, \citenamefont {Luneau}, \citenamefont {Shirman}, \citenamefont
  {Cargnello}, \citenamefont {Bare}, \citenamefont {Aizenberg}, \citenamefont
  {Friend},\ and\ \citenamefont {Frenkel}}]{timoshenko2018probing}%
  \BibitemOpen
  \bibfield  {author} {\bibinfo {author} {\bibnamefont {Timoshenko},
  \bibfnamefont {J.}}, \bibinfo {author} {\bibnamefont {Wrasman}, \bibfnamefont
  {C.~J.}}, \bibinfo {author} {\bibnamefont {Luneau}, \bibfnamefont {M.}},
  \bibinfo {author} {\bibnamefont {Shirman}, \bibfnamefont {T.}}, \bibinfo
  {author} {\bibnamefont {Cargnello}, \bibfnamefont {M.}}, \bibinfo {author}
  {\bibnamefont {Bare}, \bibfnamefont {S.~R.}}, \bibinfo {author} {\bibnamefont
  {Aizenberg}, \bibfnamefont {J.}}, \bibinfo {author} {\bibnamefont {Friend},
  \bibfnamefont {C.~M.}}, \ and\ \bibinfo {author} {\bibnamefont {Frenkel},
  \bibfnamefont {A.~I.}},\ }\href@noop {} {\bibfield  {journal} {\bibinfo
  {journal} {Nano letters}\ }\textbf {\bibinfo {volume} {19}},\ \bibinfo
  {pages} {520} (\bibinfo {year} {2018}{\natexlab{b}})}\BibitemShut {NoStop}%
\bibitem [{\citenamefont {Vaarkamp}(1998)}]{Vaarkamp1998}%
  \BibitemOpen
  \bibfield  {author} {\bibinfo {author} {\bibnamefont {Vaarkamp},
  \bibfnamefont {M.}},\ }\href@noop {} {\bibfield  {journal} {\bibinfo
  {journal} {Catalysis today}\ }\textbf {\bibinfo {volume} {39}},\ \bibinfo
  {pages} {271} (\bibinfo {year} {1998})}\BibitemShut {NoStop}%
\bibitem [{\citenamefont {Vrieze}(2012)}]{vrieze2012model}%
  \BibitemOpen
  \bibfield  {author} {\bibinfo {author} {\bibnamefont {Vrieze}, \bibfnamefont
  {S.~I.}},\ }\href@noop {} {\bibfield  {journal} {\bibinfo  {journal}
  {Psychological methods}\ }\textbf {\bibinfo {volume} {17}},\ \bibinfo {pages}
  {228} (\bibinfo {year} {2012})}\BibitemShut {NoStop}%
\end{thebibliography}%

\section*{Appendix 1: Empirical Bayes method for optimization of Bayesian parameters}
\label{App:paraopt}
To estimate the optimal prior parameters, $z$ and $z_w^0$, which are obtained from the minimization of the free energy $\mathrm{BFE}(\boldsymbol{c}, z, z_w^0)$, 
we derive the partial derivative of $\mathrm{BFE}(\boldsymbol{c}, z, z_w^0)$ with respect to $z$ and $z_w^0$.  
Then, to find the relative minima of $\mathrm{BFE}(\boldsymbol{c}, z, z_w^0)$, 
we set $\partial \mathrm{BFE}(\boldsymbol{c}, z, z_w^0) / \partial z =0$ and 
$\partial \mathrm{BFE}(\boldsymbol{c}, z, z_w^0) / \partial z_w^0 =0$ and derive the following self-consistent equations:
\begin{eqnarray}
z = 
\left( 
-\frac{\partial}{\partial z}\boldsymbol{\mu}^\mathrm{T}
\Lambda^{-1}
\boldsymbol{\mu} 
+ \boldsymbol{y}^\mathrm{T}
\frac{\partial \Sigma^{-1}}{\partial z}\boldsymbol{y}^{-1} 
+ \mathrm{Tr}\left( \Lambda\frac{\partial \Lambda^{-1}}{\partial z}\right)\right),
\label{eq:z_BFE_partial}
\end{eqnarray}
\begin{eqnarray}
z^0_{\mathrm{w}} = 
(C+1)
\left(
\boldsymbol{\mu}^{\mathrm{T}}
\boldsymbol{Z}_r(\boldsymbol{c})
\boldsymbol{\mu}
+\mathrm{Tr}(\Lambda\boldsymbol{Z}_r(\boldsymbol{c}))\right), 
\label{eq:z0w_BFE_partial}
\end{eqnarray}
where we set $L=\left(\sum_{i=1}^p\frac{1}
{1+\sigma_i^2z}\right)$ and $C\equiv|\boldsymbol{c}|$. 
We solve the above equations through fixed-point iteration. 
Specifically, starting from an initial value $z=1$, 
we compute the right side of Eq.~(\ref{eq:z_BFE_partial}) and 
update $z$, and then we substitute the updated value $z$ in Eq. (\ref{eq:z0w_BFE_partial}). 
We then repeat the operation and obtain the convergent value of $z$ and $z^0_{\mathrm{w}}$ as a solution of the above nonlinear equation. 

\section*{Appendix 2: Parameter optimization based on cross-validation error}
\label{App:CVE}
In a previous study \cite{Akai2018}, regularization parameter $\lambda$ was optimized by
using the cross-validation error (CVE), which is denoted 
$\lambda_{\mathrm{CVE\;min}}$. Cross-validation (CV) enables the learning of a regression model using limited data without overfitting. In CV, the limited data are divided into training data and testing data. The estimated regression model for the training data is evaluated on the testing data using the selected bases represented by the indicator vector $\boldsymbol{c}(\lambda)$. To eliminate the influence of bias in the data division, the training–evaluation procedure is repeated for a different division of the data. After several repeats, the results are averaged to provide the evaluation value of the parameter combination represented by the indicator vector $\boldsymbol{c}(\lambda)$. The CV that divides the data into $M$ pieces and uses a one-division evaluation is called $M$-fold CV. In the present research, the evaluation is performed by $10$-fold CV (i.e., $M = 10$). 

Simply stated, the minimizer $\lambda_{\mathrm{CVE\;min}}$ of the CVE can be regarded as the optimal $\lambda$, but $\lambda_{\mathrm{CVE\;min}}$ tends to select a variable combination that is not very sparse \cite{Igarashi2018}. 
Thus, a heuristic criterion called the "one-standard-error (1SE) rule", by which the largest 1SE that provides a larger CVE than the minimal CVE by at most the CVE's standard error is taken \cite{Murphy2012}, is used in the previous study \cite{Akai2018}. 

\section*{Appendix 3: Comparison of magnitude spectra obtained by using AdvFT and FT bases}
 \label{subsec:FTcompare}
\begin{figure}
  \centering
  \includegraphics[width=3in]{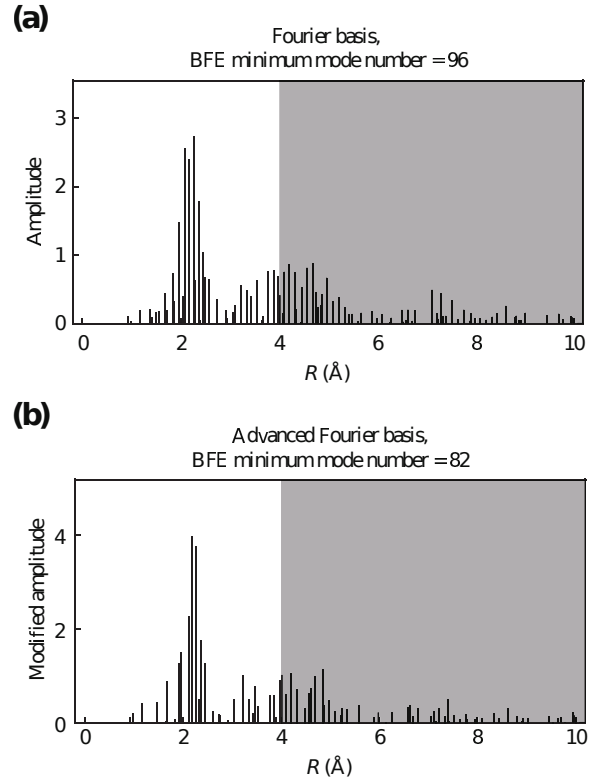}
  \caption{
Comparison between the results of the magnitude spectrum by FT and those modified by AdvFT.
(a) Magnitude spectra obtained by FT, where the $L1$-regularization parameter $\lambda$ is optimized by BFE minimization, as shown in Fig. \ref{fig:HyperParamEst-DWsigmavslambdaCVEandFEmap-slices}(a). 
(b) Magnitude spectrum using AdvFT, as shown in Fig. \ref{fig:BFEandCVE1SE_Est_mode}(f) modified by multiplying both the polynomial term $k^2/R_j^2$ and the exponential component, which is included in the FT basis, and computing the maximum value in terms of $k$ space. 
   }
 \label{fig:BFE_AMP_ByBaseFT_Compare}
\end{figure}

In this section, we compare the magnitude spectra obtained by using AdvFT and FT bases.
Fig. \ref{fig:BFE_AMP_ByBaseFT_Compare}(a) shows the magnitude spectra obtained by the BFE-minimized solution with the FT basis, where $\lambda_{\mathrm{BFEmin}}=1.9\times10^{-2}$ is selected by BFE minimization, 
indicated with a diamond mark 
in Fig. \ref{fig:HyperParamEst-DWsigmavslambdaCVEandFEmap-slices}(a). 
Although this result is significantly different from that obtained by AdvFT 
in Fig. \ref{fig:BFEandCVE1SE_Est_mode}(f), the result from AdvFT can be converted into magnitude spectra 
similar to that by FT, as presented in Fig. \ref{fig:BFE_AMP_ByBaseFT_Compare}(b). 
Since the FT basis is constructed from $\sin 2k R_j$ and $\cos 2k R_j$, 
the polynomial term $k^2/R_j^2$ and the exponential component 
are involved in the magnitude of FT. 
Additionally, such terms are embedded in the AdvFT basis as shown in Eqs. (\ref{eq:ASincoeff}) and (\ref{eq:ACoscoeff}). 
Then, by multiplying these terms in the magnitude spectra obtained by AdvFT. 
and extracting the maximum value in terms of $k$ space, 
we can obtain magnitude spectra 
that is similar to that with the magnitude of the FT, 
as presented in Fig. \ref{fig:BFE_AMP_ByBaseFT_Compare}(b). 
Thus, these magnitude spectra are consistent on the whole frame, 
which indicates that the basis function defined by large $R$ has a relatively small effect on the EXAFS signals. 
\\
EXAFS signals are considered to be insensitive to fluctuations in magnitude spectra in the large $R$ region due to the constraint on the damping effect from the finite mean free path of the photoelectron wave. Thus, the accuracy of 
the magnitude spectra at large $R$ is lower than that at small $R$. This is consistent with a commonly held view about the reliability of EXAFS data, i.e., EXAFS data analysis is limited by the mean free path of photoelectron wave propagation \cite{Stern1995,Bunker2010,Teo2012,Newville2014}. 

\end{document}